\documentclass{ws-rv961x669}
\usepackage[square]{ws-rv-van}     % numbered citation/references (default)
\usepackage{ws-rv-thm}     % comment this line when `amsthm / theorem / ntheorem` package is used
\usepackage{subfigure}     % required only when side-by-side / subfigures are used
\makeindex

\usepackage{ws-rv-van}
\makeindex
\makeindex
\begin{document}

\chapter[]{50 years of correlations with Michael Fisher \\and the renormalization group}

\author[]{Amnon Aharony}
%\index[aindx]{Author, F.} % or \aindx{Author, F.}
%\index[aindx]{Author, S.} % or \aindx{Author, S.}

\address{School of Physics and Astronomy, Tel Aviv University,\\
Ramat Aviv, Tel Aviv 6997801, Israel \\
aaharonyaa@gmail.com}

\begin{abstract}
    I start with a review of my personal and scientific interactions with Michael E. Fisher, who was my post-doc mentor in 1972-1974. I then describe several recent  renormalization group studies, which started during those years, and still raise some open issues. These include the magnets with dipole-dipole interactions, the puzzle of the bicritical points and the random field Ising model.
\end{abstract}

%\markboth{Even Page Header}{Odd Page Header} % Customized running heads

\body

%\tableofcontents

%%%%%%%%%%%%%%%%%%%%%%%%%%%%%%%%%%%%%%%%%%%%%%%%%%%%%%%%%%%%%%%%%%%%%%%%%%

\section{My life with Michael E. Fisher}\label{sec1}

\subsection{Personal}

I worked on my Ph. D. thesis with the late Yuval Ne'eman at Tel Aviv University, in parallel to my military service in the Israeli army. Expecting to be slow relative to the level of activity required in more competitive topics, I chose to work on the arrows of time, combining recent developments on the breaking of time reversal symmetry in the decay of the K-meson with the second law of thermodynamics. I thus studied both high energy physics and statistical physics, but knew very little about critical phenomena. When I wrote my thesis (while on duty on the Golan heights in September 1971), I started looking for a place for my postdoctoral training, and I almost accepted a position where I would continue my work on irreversible thermodynamics.

However, two crucial events changed my life. First, the late Joe Imry just returned from his postdoc at Cornell, and he told me about his good experience there. Second, the late Pierre Hohenberg visited Tel Aviv, and told us about Wilson's renormalization group (RG). Both  became close friends and collaborators for many years. As a result, I wrote to Michael Fisher, and applied for a postdoc position in his group. In spite of my lack of sufficient background in phase transitions, Fisher soon offered me a postdoc position. In particular, he asked me to respond within 2 weeks! Being very precise, he indicated that I shall have only two duties: do research andattend the weekly Widom-Fisher seminar. I immediately cancelled my alternative possibilities, and accepted Fisher's offer.

\begin{figure}[ht]
{\includegraphics[width=6cm]{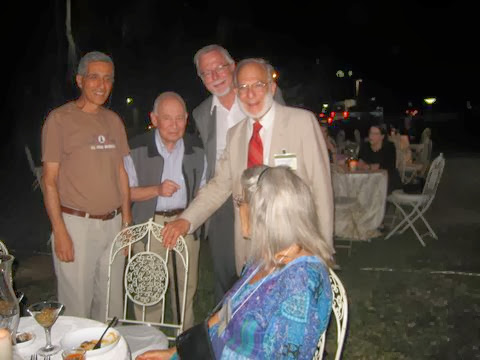}}\ \  {\includegraphics[width=6cm]{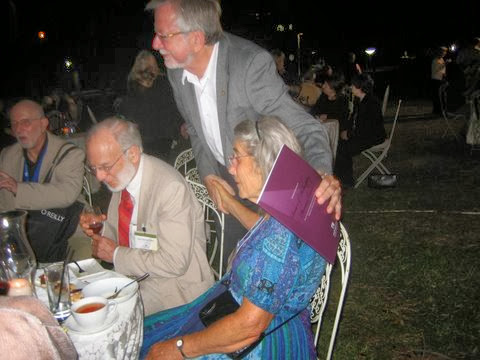}}\ \
\caption{Michael and Sorrel Fisher in 2009, when he  received an honorary PhD from the WIS. Left: with David Mukamel, Joe Imry, AA. Right: with Sir Michael Berry and AA. }
\label{fig1}
\end{figure}

My wife, my 2-year old son Ofer (now a high energy physicist at the Weizmann Institute, studying string theory -- Fisher won: he has two sons in physics...) and I used my Fulbright travel grant\footnote{Most of the postdocs at Cornell came with some external fellowships.} and arrived in Ithaca in July 1972, and were told that we missed the summer, which occurred on the day before our arrival (Ithaca is cold, and summers are short.)
We had already rented an apartment (for which Michael had kindly loaned us the first rent), and thus we quickly settled down. During the following two years, we were often invited to the Fisher home, where we enjoyed the warm hospitality of both Sorrel and Michael.
As on many later occasions, in October 1973 Michael drove me in his car to my first magnetism conference in Boston, where I talked about our joint work on  dipolar magnets (see below). We returned to Ithaca a few days before my  daughter Tamar was born (in parallel to worrying about the Yom Kippur war in Israel). Kindly, the Fishers held a party in their home, for exhibiting Tamar to the local community. I last met Sorrel in 2009, when Michael was awarded an honorary doctorate at the Weizmann Institute (Fig. 1), and I was very sad to hear about her death in 2016. It was difficult for Michael to survive without Sorrel. He reacted emotionally when I sent him the photos in Fig. \ref{fig1}.

\begin{figure}[ht]
\center{\includegraphics[width=10cm]{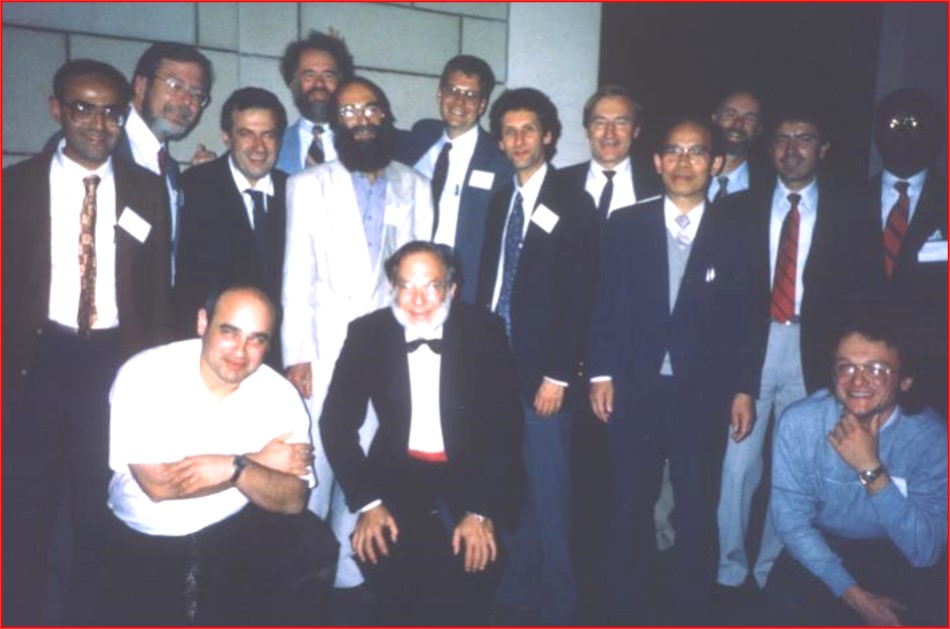}}\ \
\caption{Fisher and his students/postdocs on his 60h birthday, NAS, Washington DC 1991.  Top: Mustansir Barma, AA, Shmuel Fishman, Joseph Straley, Pierre Pfeuty, David Nelson, David Jasnow, Ubo Felderhof, Masuo Suzuki, Michael Barber, Gunduz Caginalp, Arthur Ferdinand. Bottom: Vladimir Privman, MEF, Lev Mikheev.}
\label{fig2}
\end{figure}

\begin{figure}[ht]
\vspace{-1.6cm}
\center{\includegraphics[width=15cm]{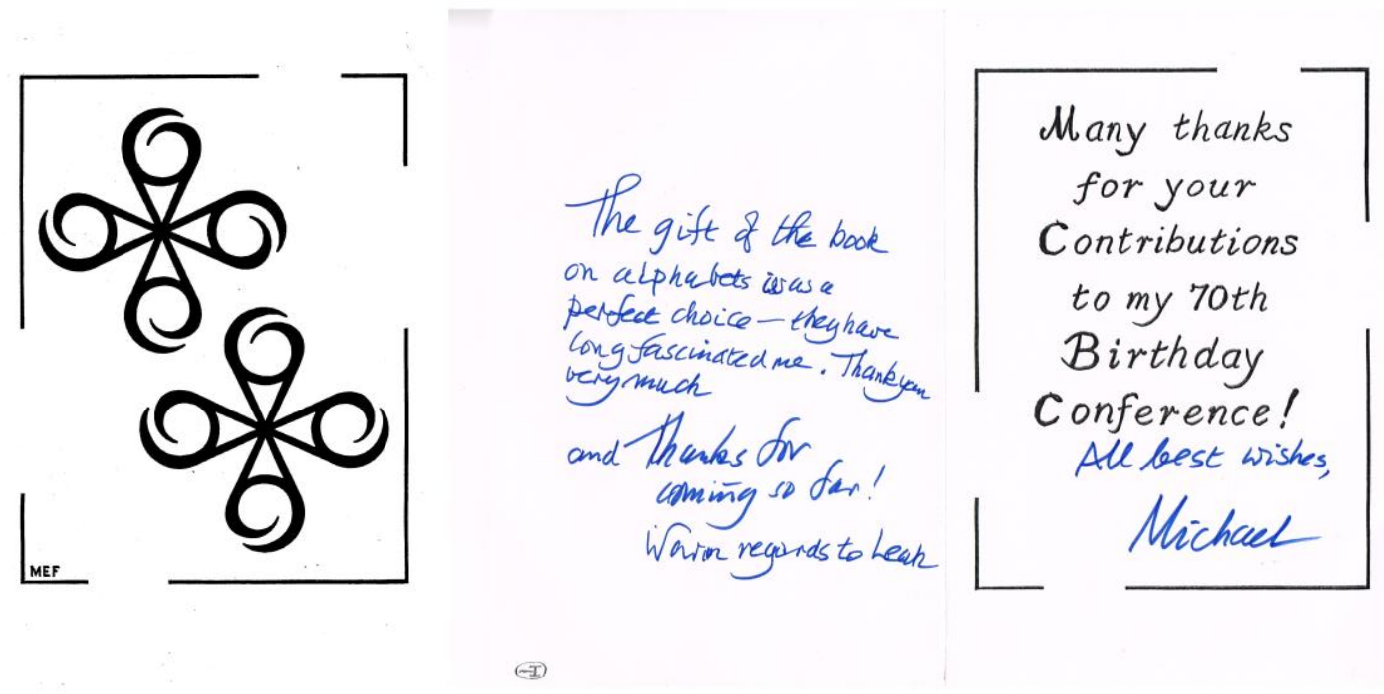}}
\vspace{-15.cm}
\caption{Card drawn and written by MEF to AA after MEF's 70th birthday. Note the typical drawings and calligraphy.}
\label{70c}
\end{figure}

\begin{figure}[ht]
{\includegraphics[width=6cm]{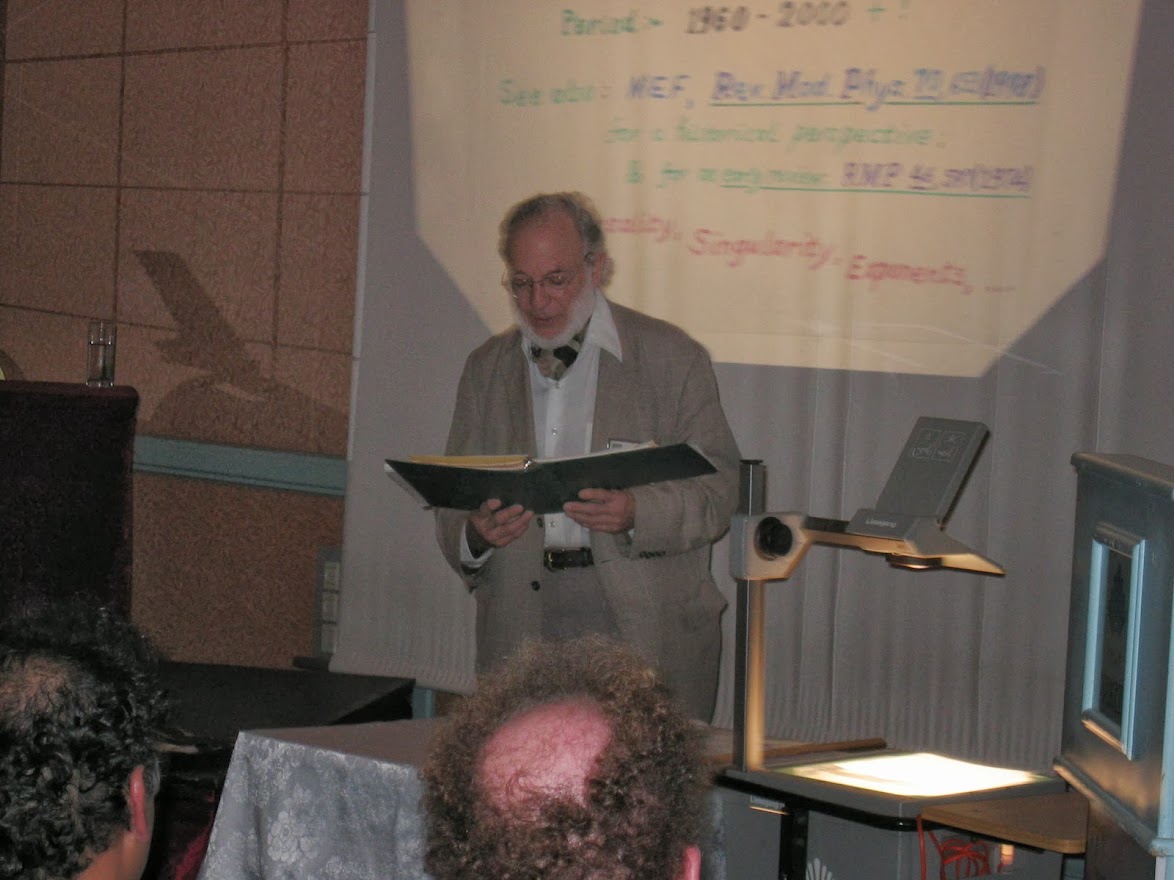}}\ \ \  {\includegraphics[width=6cm]{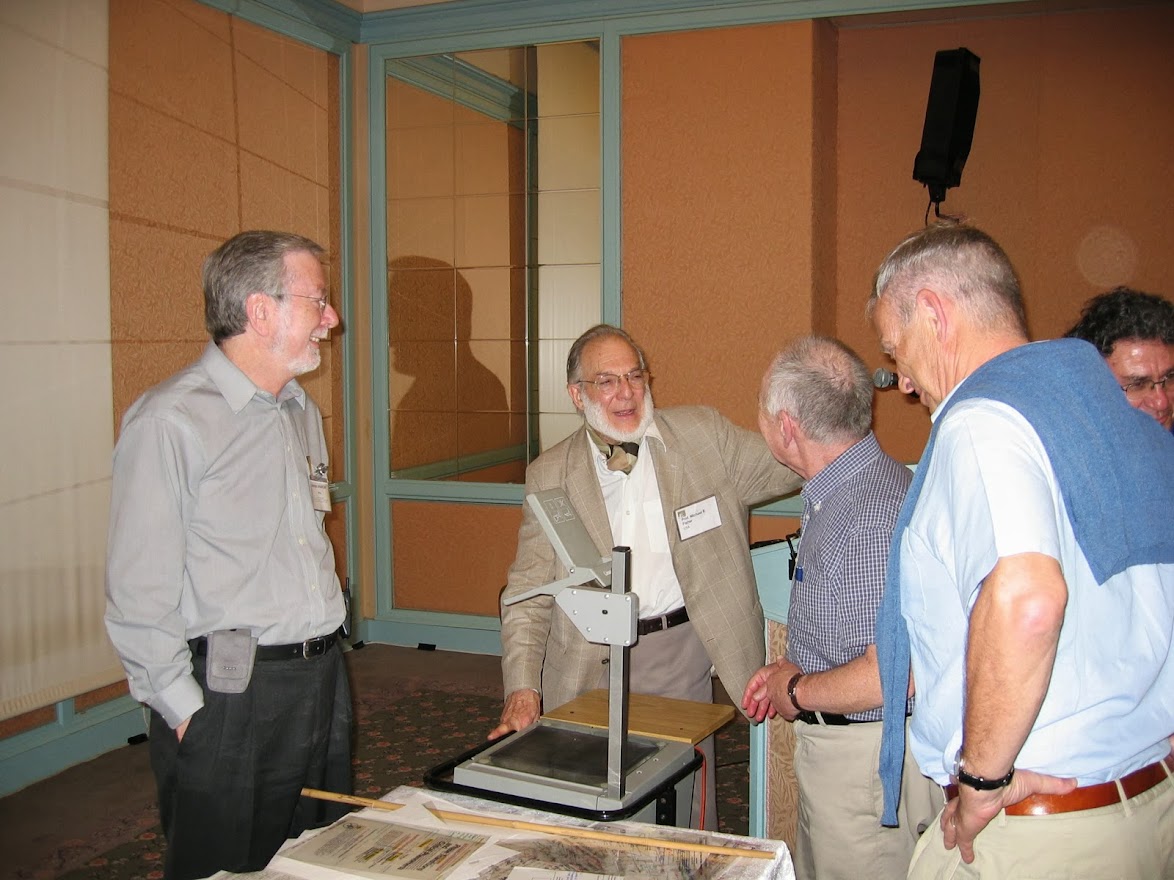}}\ \
{\includegraphics[width=6cm]{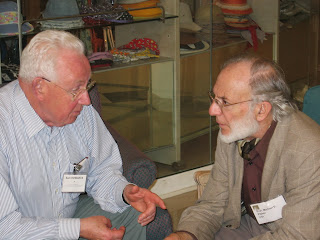}}\ \ \ \ \ \ \ {\includegraphics[width=6cm]{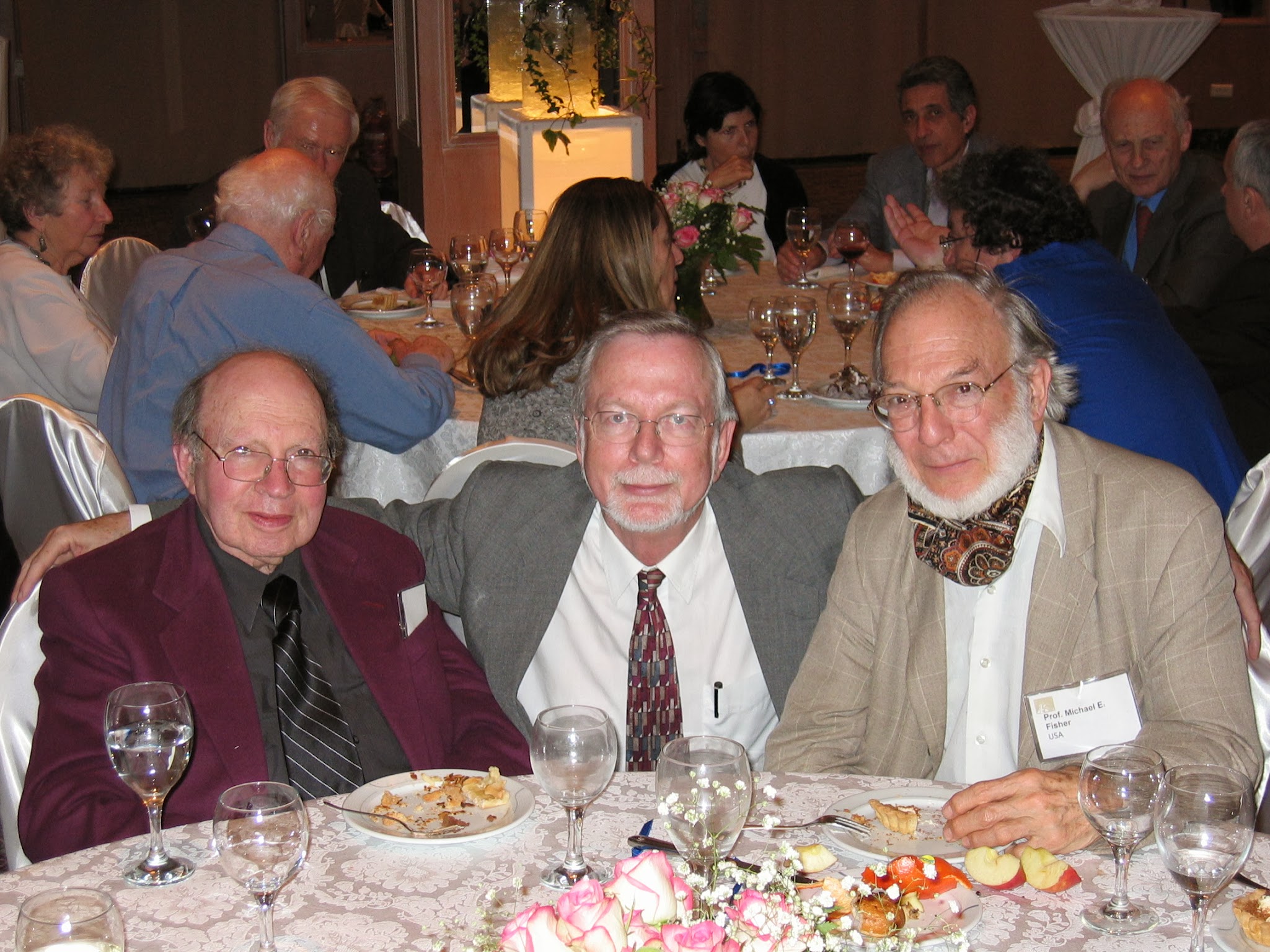}}\ \
\caption{AA's 60th birthday, Eylat, 2004. Top: Fisher lecturing with transparencies; AA, MEF, Joe Imry, \'{E}douard Br\'{e}zin and Eytan Domany.
Bottom: MEF with Sir Sam Edwards; Yuval Ne'eman, AA and MEF.  }
\label{fig3}
\end{figure}

After 1974, Michael and I met many times at conferences (e.g., Yeshiva and Rutgers Statistical Mechanics meetings,\footnote{My first Yeshiva meeting, in December 1972, was a special experience. In particular I met there my life-long friends and collaborators, the late Dietrich Stauffer and (long live) Sushanta Dattagupta, and many other players in the field.} IUPAP conferences on Statistical Physics and on Magnetism, Gordon Conferences, and more). He also visited Tel Aviv in 1980 -- on the occasion of receiving the Wolf prize\footnote{A typical Fisher story: The Wolf prize was young. Documents were prepared in Hebrew and then translated into English. In this translation, ``phase transitions" became ``transient phases", and this appeared on the  certificate given to Michael by Israel's president at the ceremony. I caught this, and asked the Wolf people to be more careful in the future. However, when they asked Michael to return the certificate for correction he refused: ``It will look much more impressive on my office wall as it is". I do not know how this ended.} (with Wilson and Kadanoff), in 1983 -- when he was honored with a Sackler distinguished visiting scholarship at Tel Aviv university,\footnote{https://ias.tau.ac.il/IAS-Distinguished-Scholars; this was before Sacklers' involvement in the opioid crisis became known; since then I no longer use the Sackler name when I refer to my School and Faculty at TAU.} \footnote{ On that visit Michael stayed for a month in an apartment with an appropriately placed desk and room for playing the guitar -- being a perfectionist in everything, Michael always insisted on proper arrangements in his rooms and offices.}, in 1988, when he participated in “Frontiers of Physics,”  Landau Memorial Conference,\footnote{“Frontiers of Physics,” Proc. Landau Memorial Conf., Eds. E. Gotsman, Y. Ne'eman and A. Voronel (Pergamon Press, Oxford, 1990).}  and in 1992 -- when he was awarded an honorary  doctorate from Tel Aviv University\footnote{http://www3.tau.ac.il/honorary$\#$phd. Michael did not bring a formal suit for the ceremony, so we rushed downtown to buy one, and I saw him wear that suit also on later occasions; even when Michael was not in Tel Aviv, Tel Aviv stayed with him...}. As I describe below, we collaborated on many topics, and I learned a lot from him --  on how to use elegant mathematics, how to write papers, how to fight with referees and with editors (especially in PRL -- both Michael and I refrained for a while from publishing as well as refereeing there) and how to present talks (with two parallel transparency projectors!). As other authors in this book describe, I also ``suffered" from his many red comments on my paper drafts, and enjoyed seeing his art of drawing figures for our papers, both of which taught me a lot. In addition, I enjoyed participating in Michael's 60th~\cite{60}, 70th~\cite{70}  and 80th~\cite{80} birthdays, see e.g. Figs. \ref{fig2},~\ref{70c}, which were great reunions of the large group of Michael's students and postdocs. Michael also attended my 60th birthday conference in Eilat, in 2004 \footnote{https://www.tau.ac.il/$~$aharony/public/eilat poster.jpg} see Fig. \ref{fig3}. In 1982, Michael Stephen, Michael Fisher and I joined others to teach a course on critical phenomena in South Africa~\cite{stellen}.\footnote{Both kindly drove on the wrong side of the road, given their prior experience.} Also, I visited Michael in Maryland in 2002. When I arrived at Dulles airport, the person at customs asked me where I am going, and I told him I am giving a colloquium on   quantum mechanics at the University of Maryland. He immediately asked when we shall get rid of airports and use  teleportation, and I told him this will not happen in my lifetime. I mentioned this conversation in my colloquium, and Michael immediately reacted: ``This must have been our student"! Note: my talk was about quantum interference in mesoscopic devices, and at that time Fisher raised questions about the relevance of quantum mechanics to condensed matter physics~\cite{MEFq}, but we did not debate on this.
Finally, I must have done something right: Michael also invited my first doctoral student, the late Shmuel Fishman, to be a postdoc in his group (first at Cornell and then at Maryland; Fig. \ref{fig2}).

\subsection{Scientific}

My time at Cornell was unique.\footnote{In a talk celebrating Wilson, David Mermin called these years "the Valhalla years of Cornell".} The Wilson-Fisher paper on the $\epsilon-$expansion~\cite{WF} had just appeared, and the authors were there to help with its applications to phase transitions in a variety of physical systems. That paper, in which they considered the dimensionality $d=4-\epsilon$ as a continuous variable, allowed simple analytic calculations (instead of the rather more complicated numerics proposed by Wilson earlier). This opened the field to many researchers, and led to the flourishing of critical phenomena ever after.\footnote{Since I came into the field in 1972, I do not feel competent to review its earlier history, accept for noting that Wilson and Fisher followed the scaling ideas set earlier by Ben Widom and Leo Kadanoff, and Widom was also there -- open for discussions and help. Fisher himself reviewed this history in Ref. \cite{MEFRMP}.} Half a century later, this is one of the main reasons to celebrate Michael Fisher and the RG in this book.

I overlapped at Cornell with many postdocs and students, both in the Baker laboratory of Chemistry, where Fisher was centered (with Ben Widom, who told me proudly that he got Fisher there, but was unable to contribute a chapter to this book), and in the Clark Hall of Physics. The list included Pierre Pfeuty and Helen Au-Yang, with whom I shared an office, David Jasnow, Howard Tarko, Alastair Bruce, Michel Droz, Michael Kosterlitz, David Nelson, Eytan Domany, Paul Gerber, Joe Sak and many more, not to mention Widom's group (e.g., Jim Bartis) and Wilson's group (e.g. H. R. Krishnamurthy). Also, Bob Griffiths was there on sabbatical. Among others, Pfeuty talked about his work with Elliot on the Ising model in a transverse field, and Kosterlitz talked about his work with Thouless on the XY model in two dimensions The atmosphere was friendly, and Michael encouraged joint discussions and collaborations.\footnote{We also had the good fortune to read the preprints of chapters from Ashcroft and Mermin's ``Solid State Physics" (I still have a pile of these heavy typed files), and be inspired by both authors.}

Ken Wilson gave a course on the RG, which he also gave in Princeton (later published by  Wilson and Kogut~\cite{WK}). Most of us worked on the RG, and often the course gave us the hints we needed for making progress in our research. Many visitors came through, and we heard their seminars both on the blackboard in the Fisher-Widom seminar and in the more conventional physics seminar. I brought a pile of slides based on my doctoral thesis (transparencies only began being used then), but did not use them -- being forced to use only the blackboard. Away from this seminar, Michael did use transparencies; he kept using them all his years, avoiding the more modern use of power point. In particular, I liked his way of using two projectors in parallel, so that one could compare theory and experiments, or keep track of definitions. I adopted this method for many years, until I did switch to power point. Several papers in this volume also mention Michael's  questions to seminar and conference speakers. Although sometimes looking sleepy, he usually understood much of the talks, and posed important questions for which the speakers were not always prepared. In the following years I tried to imitate Michael also in this respect, which may have given me the reputation of being ``brusque".

Immediately after my arrival, Michael suggested that I study the dependence of the critical temperature on the many physical parameters in specific systems.
He later returned to that problem in his work on non-linear fields, and I return to it in Sec. 2.1 below. However, shortly after that I attended a physics seminar in Clark Hall, discussing some aspects of dipole-dipole interactions. Already during the talk I realized that the propagator related to the two-spin correlation function of such interactions is similar to the propagator of the photon (which vanishes for longitudinal fluctuations), which I knew from my field theory courses at Tel Aviv. This propagator projects out longitudinal correlations. I immediately spoke to Michael about this, and we agreed that the critical behavior of dipolar systems is a better direction to follow. Indeed, this project soon ended up with six papers~\cite{1,2,3,4,5,6}, with the first appearing  already in March 1973. However, only two of these were co-authored by Michael! The `Heisenberg' dipolar interactions involve scalar products like $({\bf S}({\bf R}^{}_i)\cdot{\bf R}^{}_{ij})$, involving both the coordinate vector ${\bf R}^{}_{ij}$ and the spin vector ${\bf S}({\bf R}^{}_i)$, and therefore we used $n=d=4-\epsilon$, where $n$ is the number of spin components.\footnote{The equality $n=d$ arises in many physical systems. Generally, one can extrapolate to the physical limit $n=d=3$ along many trajectories, e.g., at fixed $n=3$ and expand in $\epsilon=4-d$, expand  in $\epsilon$ along $n=d=4-\epsilon$, at fixed $d=3$ and expand in $1/n$, etc. Similar questions arise for $n=d=2+\epsilon$ and $n=d=1+\epsilon$. At the end, all these expansions and extrapolations should approach the same limit at $n=d=3$. It will be useful to compare all these with existing modern tools.}  These scalar products also required angular integrations in continuous $d$ dimensions. This connection between space and spin also led me to generalize the Wilson-Fisher paper (which had $n=2$ coupled Ising spins) to systems with cubic symmetry, i.e., coupled $n$ component spins~\cite{cubic,AADG}, which raised issues of competing fixed points (FPs), still being investigated today (see below).  After we published these two papers, Michael allowed me to proceed on my own, taking advantage of Wilson's course and including Wilson's direct diagrammatic way of calculation~\cite{3}, antiferromagnets~\cite{4}, anisotropies (in which $n<d$)~\cite{5}, $1/n-$expansions (in which $n$ was also treated as a continuous variable) and more. Still at Cornell, I also wrote papers on dipolar systems with Alastair Bruce~\cite{dipEOS}. Needless to say, all my papers from Cornell acknowledge both Fisher and Wilson for their advice.

\begin{figure}[ht]
\center{\includegraphics[width=12.5cm]{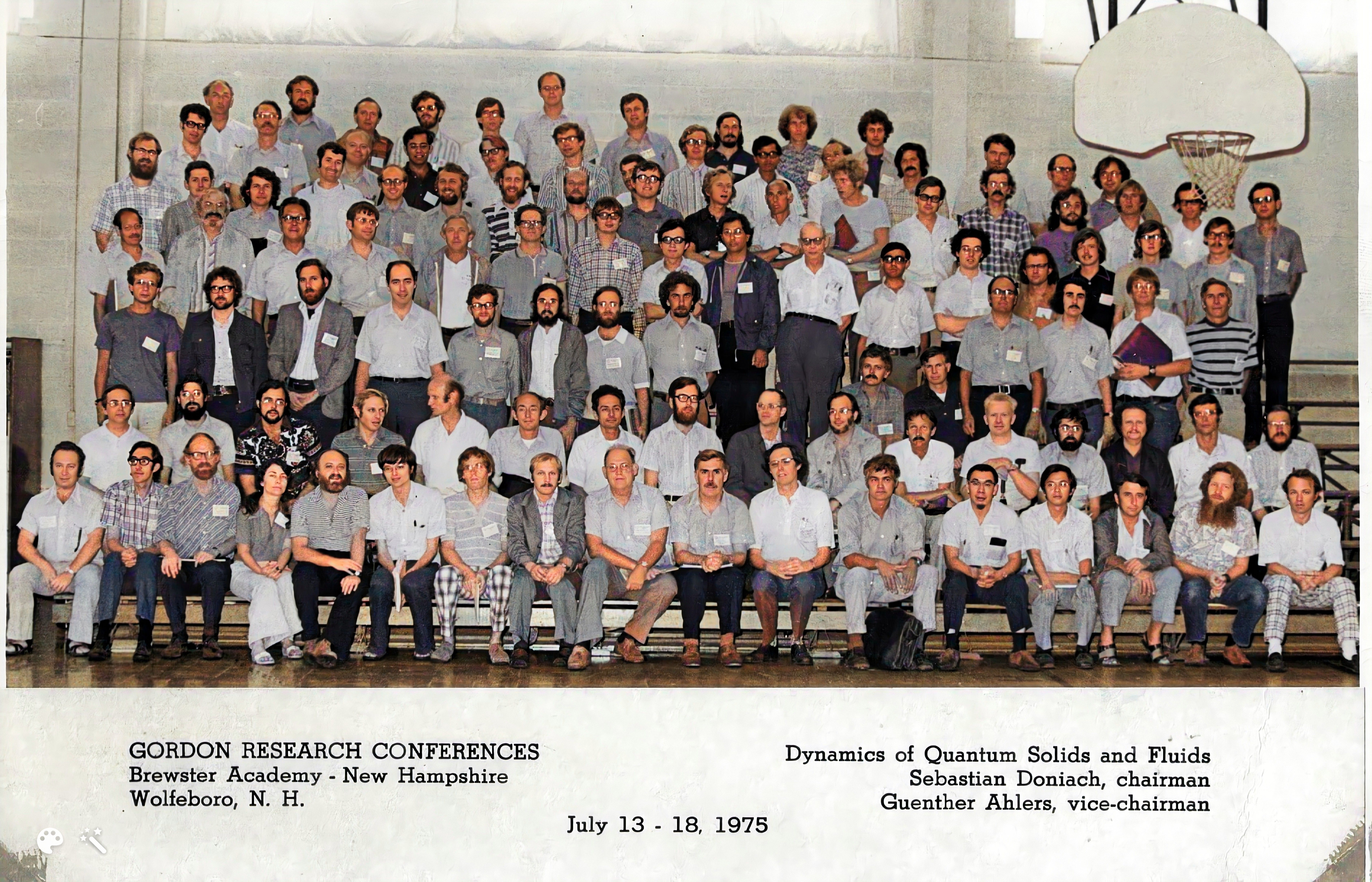}}
\caption{Color version of the original black and white photo of the participants in the Gordon conference on Dynamics of quantum solids and fluids:  cooperative phenomena in solids and fluids, July 1965. David Nelson, Bob Birgeneau and AA so far identified Aharony, Ahlers, Als Nielsen, Bak, Baker, Berker, Birgeneau, Betts, Bishop, Bruce, Chang, Ditzian (now Kadanoff), Doniach, Emery, Fisher, Grest, Greytak, Griffiths, Hertz, Hohenberg, Imry, Kadanoff, Kitchens, Kosterlitz, Krishnamurthy,  Luther, Martin, Mandelbrot, Nelson, Nickel, Passell, Pindak, Pynn, Rice, Rohrer, Rudnick, Siggia, Stephen, van Leeuwen, Widom, Wolf and Wortis. Many of these are also mentioned in the text. I leave it for the readers to connect names with pictures.}
\label{gordon}
\end{figure}

In particular, in Ref.~\cite{6} I was able to use the Fisher-Wilson approach to show that the upper critical dimension for the uniaxial dipolar problem ($n=1$) is 3, with logarithmic corrections,\cite{6} confirming the earlier Larkin-Khmelnitzkii calculations (Larkin then visited Cornell, before he knew English, and Khmelnitzkii later became a friend) and correcting the simple Landau theory approach (without the logarithmic corrections) used earlier for structural phase transitions (about which I learned from Alastair Bruce, who was a student of Roger Cowley). Guenter Ahlers (whom I also met through Michael, and later collaborated with~\cite{ahlers}) later received at a Gordon conference (Fig. \ref{gordon}~\footnote{David Nelson reminds me that the conference was announced in Science magazine as ``solids and fruits", to which Seb Doniach proposed an improvement: ``salads and fruits".}) a prize of 1/3 of a log (of wood) for confirming these corrections experimentally~\cite{GAlog}.  Unfortunately, Guenter was unable to contribute to this book. I return to dipolar systems below, in Sec. 2.1.

During my years at Cornell, many experimentalists were attracted to visit and understand the implications of the RG on their work. I later enjoyed very much collaborations with many of these experimentalists, including, e.g., Ahlers, Birgeneau, Cowley, Garland, Litster, and Rohrer. One special example was the visit of Alex M\"{u}ller, who worked on structural phase transitions and on mulicritical points in perovskites. This visit resulted in many collaborations with Alex and with Alastair Bruce~\cite{AB,KAM,bruce2}.  Fortunately, after a visit in Tel Aviv Alex ignored my enthusiastic recommendation to stay with structural transitions, and he switched to working on high temperature superconductivity, for which he was awarded the Nobel prize.
Similarly, Heini Rohrer ignored my advice to stay in magnetism (we collaborated on the random field multicritical antiferromagnets, see below), and went on to receive the Nobel prize for the scanning tunneling microscope. Apparently, doing experiments on phase transitions and talking to theorists are good preparations for Nobel prizes... Multicritical points were also studied then by  Kosterlitz, Nelson, and Fisher~\cite{KNF}, and later by Domany, Mukamel and Fisher~\cite{DMF}. Both of these papers contain examples of Fisher's special talent for drawing figures. Both Domany and Mukamel later became friends and collaborators. I return to this topic below, in Sec. 2.2.

In May 1973, less than a year after my arrival at Cornell, Mel Green and Jim Gunton organized a meeting at Temple University, entitled ``Renormalization Group in Critical Phenomena and Quantum Field Theory". Kindly, Fisher took me along to that meeting, and gave me a chance to meet all the great people who were working on that title. I talked on the dipolar systems, and met Paul Martin (who hosted me at Harvard in the fall of 1974, where I learned about random systems from Alan Luther and Geoff Grinstein), Shang-keng Ma (who hosted me at UCSD in the spring of 1975, starting a long collaboration on random systems), and Bert Halperin and Pierre Hohenberg (who hosted me at Bell Labs in the summer of 1975, starting a long collaboration on universal amplitude ratios and starting my long collaboration with Bob Birgeneau). I also met some of the experimentalists mentioned above, and people like Franz Wegner, Eberhard Riedel and \'{E}douard Br\'{e}zin. In the following years, all of these people worked on the same topics as myself, we corresponded a lot, and became good friends. The `RG community' of those early years was stimulating and friendly, and I have been grateful to Michael for helping me become part of it. Another result of that meeting was that Green invited me to write Ref.~\cite{AADG}, joining a select group of the above list.\footnote{That series of 20 books, first edited by Cyril Domb (Michael's mentor at King's college) and Green and then by Domb and Joel Lebowitz (so far leader of 124 Yeshiva$\rightarrow$ Rutgers meetings), contains many great reviews on phase transitions and critical phenomena.}

As I mentioned, Wilson also gave the same course at Princeton, where it was attended by \'{E}douard Br\'{e}zin and David Wallace. The three of them then used the RG to derive the universal scaling function of the equation of state~\cite{brezin}.  (Br\'{e}zin and I continued to read and write about each other's papers for many years).  As soon as we saw that paper,  Michael and I realized that similar universal scaling function should also exist for the two-spin correlation function. This resulted in the next two joint papers on the $\epsilon-$expansion of this function~\cite{sca1,sca2}, and by my own paper on its $1/n-$expansion~\cite{katanin}. The $\epsilon-$expansion demonstrated  for me Michael's excellent mathematical insights. In Fourier space, this function depends on the `momentum' $q$, and the perturbative expansion (in the quartic spin terms) only contributed to this $q-$dependence at second order, apparently yielding contributions only at second order in $\epsilon$. However, the short range correlations were known to have terms which behave like the energy~\cite{FL}, involving, e.g., the specific heat exponent $\alpha$, which starts at order $\epsilon$. Michael soon realized that the terms of first order in $\epsilon$ cancel out, so that our expansion of the correlation function fully agreed with the known scaling forms, at both short and long distances. This work, and Michael's insights into the universality of scaling functions, later led me to work on universal amplitude ratios~\cite{privman}. In fact, these topics came up again during Michael's visits to Tel Aviv, in 1980 and 1983, resulting with two papers~\cite{AF1,AF2} on analytic and non-analytic corrections to scaling.\footnote{After this paper appeared on the arXiv, Jacques Perk kindly reminded me that higher order series showed that these papers lacked higher order log terms (which could not be seen by the short series we had at the time).~\cite{perk}} In these papers we may have accomplished Michael's first assignment to me, back in 1972.

I cannot end this historical introduction without mentioning another `interaction' with Michael. In 1989, he spoke at the Gibbs symposium, and emphasized the Gibbs rule on the convexity of geometrical surfaces in thermodynamic surfaces. The summary of that talk~\cite{gibbs} ends with the following: ``The striking phase diagrams observed in doped crystals displaying high-$T^{}_c$ superconductivity [12] suggest the intriguing Gibbsian question: What, if anything, replaces the convexity of the $U(S,V)$ surface when there is a variable density of frozen or quenched random impurities?" Ref. [12] refers to a paper~\cite{A5}, in which some of us tried to predict details of the phase diagrams of the recently discovered high-temperature supercoductors by M\"{u}ller and Bednorz (mentioned above). That phase diagram was drawn only schematically, and some lines apparently did not seem to obey the Gibbs rules. Michael noticed that, and he would not miss any occasion to tell me that he keeps watching me, following the biblical rule ``spare the rod -- spoil the child"~ \cite{bibl}. OK, Michael, I am proud to have been your scientific child!

%%%%%%%%%%%%%%%%%%%%%%%%%%%%%%%%%%%%%%%%%%%%%%%%%%%%%%%%%%%%%%%%%%%%%%%%%%

\section{Recent developments}

 The wave of econophysics, sociophysics and the like `pushed' me away from the `modern' statistical physics community. Therefore, my interests concentrated on random systems, magnetism, quantum mesoscopics and spintronics. However, many statistical physicists stayed within real physics, and I kept following the literature on critical phenomena. I often returned to that field,  and wrote on  topics which in many respects followed Fisher's legacy and/or addressed issues which remained open. Examples include Ref.~\cite{AH}, related to Fisher's interest in random systems~\cite{singh,gibbs}, and Ref.~\cite{mef70},  related to multicritical points~\cite{KNF}.

Below I mention a few issues which I think are still open, even after 50 years. Celebrating the 50th anniversary of the $\epsilon-$expansion, it may be appropriate to remember that this expansion treated the dimensionality of space as an algebraic abstract variable, without any geometrical picture of a real system with 3.99 dimensions. One attempt to find such systems came up when I met Benoit Mandelbrot at Harvard in 1980. Mandelbrot's fractals are geometrical constructions with non-integer dimensions, which are not translationally invariant but in which the `mass' $M$ within a volume of linear size $L$ scales as $M\sim L^{D}_{}$, identifying $D$ as the (non-integer) fractal dimensionality. It turned out that real space RG, in the original spirit of Kadanoff, becomes exact on many fractals,~\cite{gefen1}, and the ease of such calculations started a large wave of publications. However, it turned out that the physics of fractals required much more than just the fractal dimensionality,\cite{gefen2}, and this path of research did not find a geometrical description of the Wilson-Fisher non-integer $d$. However, fractal models have turned out to be very useful to model non-translationally invariant physical systems, e.g., percolation clusters. Hence my long personal involvement with percolation~\cite{book}.\footnote{In fact, Michael worked on percolation clusters much before I met him~\cite{MEFperc}. He also discussed fractal vesicles in a conference which I organized with Jens Feder~\cite{MEFfractals}.}

A major development in recent years was the entrance of field theorists into the field of critical phenomena, many years after Wilson set the example (not to mention my own background in high energy). They not only rediscovered many of the physical systems studied in the 1970'ies, but they also developed and applied many novel and very accurate methods to study them, including re-summed high-order $\epsilon-$expansions,~\cite{6eps} fixed-dimensions RG calculations~\cite{fixed-d}, conformal bootstrap calculations~\cite{boot}, and accurate Monte Carlo simulations~\cite{MC}. (To save space, I give only recent references, exemplifying the large current activity of field theorists on critical phenomena).
Specifically, these calculations identified unambiguously the stable FPs of the RG, and gave consistent and accurate values for the critical exponents, both at critical and at multicritical points. The availability of such tools (and the `pushing' of Slava Rychkov) led me recently back to  the puzzle of the bicritical points~\cite{AEK1,AEK2,AE} discussed in more detail below.\footnote{Assuming that Michael would have liked this recent work, I tried to communicate it to him, but it was too late. His E-mail address, xpectnil@umd.edu, no longer responded. (The name of this address reflected Michael's attitude to E-mail, but he did use it more after he retired. More interestingly, Michael kept copies of every letter he ever sent or received. I am sure a historian of science will find many jewels in that archive. I am told that UMD is keeping it.)} Below I also present selected other topics, in which Fisher was interested already 50 years ago, and demonstrate that even half a century later, the RG is  still an active and exciting field.

\subsection{The dipole-dipole interactions}

Fifty years after my early papers on the dipolar systems, Refs.~\cite{1,2,3,4,5,6}, sometimes people ask me if it was my father that wrote them. Well, it was me, and I am still alive and kicking. (The issue may become more tricky in 30 years, when they will ask my son Ofer the same question...).
Occasionally, some of these results are rediscovered ``independently", and I send E-mails to those rediscoverers.

In any case, every material with magnetic ions has dipole-dipole interactions, and therefore it is always necessary to check their relevance to experimental measurements. Originally, the $\epsilon-$expansion for the $n=d=4-\epsilon$ isotropic dipolar case~\cite{1,2} seemed numerically close to the short-range Heisenberg case ($2\nu^{}_{dip}=1+9\epsilon/34+\dots$ versus $2\nu^{}_{iso}=1+\epsilon/4+\dots$). Similar closeness arose for $1<<d$.~\cite{5} However, these numbers should be improved, and I still await extensions to higher order expansions and other accurate high energy methods, like those mentioned above.  (As Slava Rychkov tells me~\cite{Rychkov}, the dipolar model is not conformal, hence bootstrap is not applicable. Other methods should still work.)

As mentioned, the uniaxial Ising case is particularly interesting. The upper critical dimension is $d^{}_u=3$, and one could prove analytically that at this dimension, e.g., the susceptibility diverges near criticality as $\chi\sim |t|^{-1}|\log|t||^{1/3}$, where $t$ is the scaling field related to the temperature distance from the critical point (this is the one third of a log mentioned above), similar to the behavior of the short range Ising model at $d=4$. However, the next correction, of order $\log|\log|t||/\log|t|$, has different amplitudes in these two cases~\cite{brezin2}. Note: this dimensionality shift from $d$ up to $d+1$ dimensions differs from that of the random field shift from $d$ to $d+2$ dimensions, discussed below~\cite{AIM}; the latter results from supersymmetry~\cite{parisi}, and therefore {\it all} the terms in the $\epsilon-$expansions map from $d$ to $d+2$.  Interestingly, unlike at $d<d^{dip}_u$, universal amplitude ratios {\it at} $d=d^{dip}_u$  have logarithmic corrections, e.g. the free energy per correlation volume~\cite{amp}
\begin{align}
\xi^2_{}\xi^{}_\parallel C t^2/k^{}_B=(3/32\pi)|\log|t||,
\end{align}
where $C$ is the singular specific heat, $\xi$ is the correlation length, $\xi^{}_\parallel=g_d^{1/2}\xi^2_{}$, $g^{}_d$ measures the strength of the dipolar interactions and $k^{}_B$ is the Boltzmann factor. This implies logarithmic corrections to the hyperscaling relation, $d\nu=2-\alpha$, which do not arise at $d<3$, where the free energy per correlation volume is a universal temperature-independent constant.~\cite{privman}
\footnote{To show the close community in those days, the acknowledgements in that paper include G. Ahlers, J. Als-Nielsen, R. Birgeneau, P. C. Hohenberg, J. M. Kosterlitz, D. R. Nelson and E. Siggia. Unfortunately the first 3 experimentalists were unable to contribute to this volume.}

The situation is even more interesting when randomness is included. In that case the expansion in $\epsilon=3-d$ starts at order $\epsilon^{1/2}$, and the logarithmic corrections for the susceptibility at $d=d^{}_c=3$ are replaced by $\chi\sim |t|^{-1}\exp[(D|\log|t||)^{1/2}]$, with the universal $D=9/[8\log(4/3)+53]$, different form $D=6/53$ for the short range Ising model in $d=4$~\cite{new}. I found one reference that claims to have seen a crossover to this behavior~\cite{expdd}, but the evidence there is somewhat indirect.

My recent  interest in the dipolar system arose when my colleague at Ben Gurion University Moshe Schechter told me about his paper on LiHo$^{}_{x}$Y$^{}_{1-x}$F$^{}_4$~\cite{shechter}.\footnote{This seems to be a relative of LiTbF$^{}_4$, where Ahlers originally measure the 1/3 log correction~\cite{GAlog}.} This is a dilute uniaxially anisotropic quantum dipolar ferromagnet. At $x=1$, an external transverse field $B$ on this non-random ferromagnet causes a decrease of the critical temperature from $T^{}_c(B=0)\approx 1.53$K all the way down to a zero temperature quantum transition at a field of $B\approx 4.9$T. $T^{}_c$ also decreases with decreasing $x$ (the Y ion is not magnetic), approaching a spin glass phase at low $x$. Shechter {\it et al.} review various numerical calculations of the dependence of $T^{}_c$ on both $B$ and $x$, including their own -- which emphasized the quantum fluctuations due to off-diagonal spin terms which come from the dipolar interactions.

Since fluctuations are the ``bread and butter" of the RG, I now propose to use the  RG for an alternative approach to this system. %Since the main variation of $T^{}_c$ with the magnetic field $B$ andwith the concentration $x$ is at high temperatures, I ignore the quantum fluctuations.
As I show, LiHo$^{}_{x}$Y$^{}_{1-x}$F$^{}_4$ practically contains all the topics studied by RG during the recent decades: crossover from short range interactions to isotropic dipolar interactions, then to anisotropic (uniaxial) dipolar interactions, then to random (dilute) uniaxial Ising behavior (both random exchange and random fields), and finally adding quantum effects. The order of these crossovers may change, depending on the relative strengths of the relevant parameters. Below I extend many of my old papers on some of these topics, and suggest how one can derive the critical temperature as function of the short/long interaction strengths, the spin anisotropy, the dilution and the low temperature quantum effects. This discussion is rather technical, and the unmotivated reader can skip it.

 At first, I ignore quantum fluctuations. These can be added later, by adding another dimension of imaginary times of a width that grows with decreasing temperature~\cite{elliot,suzuki,sachdev,Qshift}.
Classically, we start with the isotropic Heisenberg short range system. Pure LiTbF$^{}_4$ then requires adding the isotropic dipole-dipole interactions, with strength $g^{}_d$, as treated in Ref. \cite{2}, and the spin anisotropy, with strength $g^{}_a$, as treated in Ref. \cite{5}.
We start at $B=0$ and $x=0$. The classical (rescaled)  Ginzburg-Landau-Wilson Hamiltonian of the $n-$component spin model in momentum space is\footnote{Here and below we denote $\bar{\cal H}=-{\cal H}/(k^{}_BT)-\sum^{}_{\bf R}\big[|{\bf S}^{}_{\bf R}|^2/2+u[|{\bf S}^{}_{\bf R}|^4\big]$.~\cite{AADG}}
\begin{align}
\bar{\cal H}^{}_0(\{{\bf S}\})=-\sum^{}_{\alpha\beta}\Big[\frac{1}{2}\int_{\bf q}U^{\alpha\beta}_2({\bf q})S^\alpha_{\bf q}S^\beta_{-{\bf q}}+u^{}_{\alpha\beta}\int_{\bf q}\int_{\bf q'}\int^{}_{\bf q''}S^\alpha_{\bf q}S^\alpha_{\bf q'}S^\beta_{\bf q''}S^\beta_{\bf -q-q'-q''}\Big],
\label{H0}
\end{align}
where the integrals are over a spherical Brillouin zone of radius 1, and~\cite{2}
\begin{align}
U^{\alpha\beta}_2({\bf q})&=[r^{}_\alpha+q^2]\delta^{}_{\alpha\beta}+g^{}_d q^\alpha q^\beta/q^2,\ \ {\bf q}\ne 0,\nonumber\\
&=r^{}_\alpha\delta_{\alpha\beta}+g^{}_d{\cal D}^{\alpha\beta}_{}, \ \ {\bf q}= 0,
\label{Uab}
\end{align}
with
$r^{}_\alpha=(T-T^{MF}_\alpha)/T^{MF}_\alpha$, and ${\cal D}^{\alpha\beta}$ being the (shape-dependent) demagnetization tensor. The quartic coefficients $u^{}_{\alpha\beta}$ will be chosen below (for the simplest isotropic case, $u^{}_{\alpha\beta}\equiv u$.) Our final aim is to find the shift of the critical temperature $T^{}_c$ from its mean field value, $T^{MF}_\alpha$, for the highest `initial' $T^{MF}_\alpha$, which we assume arises for $\alpha=n=\parallel$. For simplicity we ignored in $U^{\alpha\beta}_2$ the terms $q^\alpha q^\beta$ and $(q^\alpha)^2$~\cite{2,nat}.
In pure LiHoF$^{}_4$ there is also a uniaxial anisotropy, $g^{}_a=r^{}_\perp-r^{}_\parallel$,  so that
\begin{align}
r^{}_n=r^{}_\parallel=r-(1-1/n)g^{}_a,\ \ r^{}_\perp=r^{}_1=\dots=r^{}_{n-1}=r+g^{}_a/n.
\label{rra}
\end{align}
Adding a magnetic field will add more shifts in the $r$'s, see below.

 Near the isotropic short range FP, the RG iterations for the three main scaling fields give $t(\ell)=t(0)e^{\ell/\nu},\ g^{}_d(\ell)=g^{}_d(0)e^{(2-\eta)\ell},\ g^{}_a(\ell)e^{\ell\phi/\nu}$. Here, $t$ is the true scaling field for the temperature, which vanishes at the critical temperature $T^{}_c$, and $\phi$ is the crossover exponent for the anisotropy.~\cite{pfeuty}
 Eventually, for $g^{}_d(\ell)$ and $g^{}_a(\ell)$ sufficiently large, the system will cross over to the uniaxial dipolar problem, treated in Ref. \cite{6}.
 At that point, the scaling field $t(\ell)$, associated the with the single critical temperature parameter $r^{}_\parallel$,  will be proportional to the true $T-T^{}_c$, and therefore it will yield the temperature shift $T^{MF}_c-T_c$, as measured experimentally.
 %The crossover is straightforward if one of the $g$'s is much larger that the other (see below), but otherwise the calculation is complicated but doable.\cite{reid,kaul}

The $\epsilon-$expansion is based on iteratively integrating out of the partition function the $S^{}_\alpha({\bf q})$'s in the spherical shell $1-\delta\ell<|{\bf q}|<1$ and then rescaling the spin and momenta so that the momentum cutoff and the coefficient of $q^2$ return to 1.~\cite{WK}
At $d=4-\epsilon$ the recursion relation for $r^{}_\alpha$ is~\cite{pfeuty}
\begin{align}
dr^{}_\alpha/d\ell=2r^{}_\alpha+12u^{}_{\alpha\alpha}A^{\alpha\alpha}+4\sum^{}_{\alpha\ne\beta}u^{}_{\alpha\beta}A^{\beta\beta}+\dots,
\label{rrr}
\end{align}
where the dots indicate higher orders in the $u$'s,
\begin{align}
A^{\alpha\beta}=\int^>_{} G^{\alpha\beta}_{}({\bf q})/(\delta\ell),
\end{align}
the integrals are over a spherical shell in momentum space, with $1-\delta\ell<|{\bf q}|<1$ and the matrix $G\equiv [U^{}_2]^{-1}$ (for ${\bf q}\ne 0$) is~\cite{5}
\begin{align}
G_{}^{\alpha\beta}({\bf q})=\frac{1}{r^{}_\alpha+q^2}\Big[\delta^{}_{\alpha\beta}-\frac{g^{}_d}{q^2+g^{}_d Q^2}\frac{q^\alpha q^\beta}{r^{}_\beta+q^2}\Big],
\label{Gd1}
\end{align}
with $Q^2=\sum^{}_\gamma (q^\gamma_{})^2/[r^{}_\gamma+q^2]$.
For $g^{}_d=0$ this reduces to $A^{\alpha\beta}=K_d\delta^{}_{\alpha\beta}/(r^{}_\alpha+1)$, where $K^{}_d$ is the area of a $d-$dimensional sphere. In the dipolar case $G_{}^{\alpha\alpha}$ depends on the direction of ${\bf q}$, and the integrals, which depend on $g^{}_d$, are more complex~\cite{2,5,6}.
%For $g^{}_a\gg 1$ (\ref{Gd}) reduces to the uniaxial dipolar limit,~\cite{6}
%\begin{align}
%G_{}^{\alpha\beta}=\delta^{}_{\alpha,n}\delta^{}_{\beta,n}\frac{1}{r^{}_n+1+g^{}_d\cos^2\theta},
%\label{theta}
%\end{align}
%where $\theta$ is the angle between the unit vector $\hat{\bf q}$ and the $n-$axis.

Since  $g^{}_d(\ell)=g^{}_d(0)e^{(2-\eta)\ell}$ is exact [and we ignore the $\ell-$depedence of $\eta=O(\epsilon^2)$], we can substitute this $\ell-$dependence into Eqs. (\ref{rrr}) and find $r^{}_\parallel(\ell)$ and $r^{}_\perp(\ell)$. Now we can
 follow the example of Nelson and Domany:~\cite{ND} iterate until $r^{}_\perp(\ell^{}_1)=1$, obtain $r^{}_\parallel(\ell^{}_1)$, integrate the transverse spins out of the partition function and then proceed with the recursion relations for the uniaxial case with only $S^\parallel$. Note: the propagator at that stage may still have the form
 \begin{align}
G_{}^{\parallel\parallel}=\frac{1}{r^{}_\parallel+q^2}\Big[1-\frac{g^{}_d}{q^2+g^{}_d Q^2}\frac{(q^\parallel)^2}{r^{}_\parallel+q^2}\Big],
\label{Gd}
\end{align}
with $Q^2=(q^\parallel_{})^2/[r^{}_\parallel+q^2]$. It will approach
\begin{align}
G_{}^{\alpha\beta}=\delta^{}_{\alpha,n}\delta^{}_{\beta,n}\frac{1}{r^{}_n+q^2+g^{}_d\cos^2\theta},
\label{theta}
\end{align}
where $\theta$ is the angle between the unit vector $\hat{\bf q}$ and the $n-$axis, only when $g^{}_d(\ell)\gg 1$.
Although some  calculations of the crossover in the presence of both $g^{}_a$ and $g^{}_d$ was done previously~\cite{reid,kaul}, I think the above prescription is more straight forward.

Adding a  magnetic
 field $B$ (in energy units -- including the factor $g\mu^{}_B$)  on the spin component $S^{1}_{}$, adds a term $--BS^{1}_{q=0}$ to ${\cal H}^{}_0$. It is then convenient to replace $S^{1}_{\bf q}\Rightarrow S^{1}_{\bf q}+M\delta({\bf q})$. This shifts the coefficients in $U^{\alpha\beta}_2$~\cite{dipEOS},
 \begin{align}
&r^{}_1\rightarrow \bar{r}^{}_1=r^{}_\perp+12 uM^2=\bar{r}^{}_\perp+8uM^2,\nonumber\\
&r^{}_\perp\rightarrow \bar{r}^{}_\perp=r^{}_\perp+4 uM^2,\ \ {\rm applies~ for}\  \beta=2,\dots,n-1,\nonumber\\
&r^{}_n\rightarrow \bar{r}^{}_\parallel=r^{}_\parallel+4 uM^2,
\label{r123}
\end{align}
where, for simplicity, we set $u^{}_{\alpha\beta}\equiv u$. The value of $M$ is determined by requiring that the new coefficient of $S^1$ vanishes~\cite{dipEOS},
\begin{align}
(r^{}_\perp+g^{}_d{\cal D}^{11}+4u M^2)M=B,
\end{align}
and at small $M$ we can use $M\approx B/(r^{}_\perp+g^{}_d{\cal D}^{11})$. In addition, this shift also generates a cubic spin term, $4u M \sum^{}_{\alpha}S^1_{\bf q}S^\alpha_{\bf q'}S^\alpha_{\bf -q-q'}$.~\cite{dipEOS}

Equations (\ref{r123}) imply
\begin{align}
\bar{r}^{}_1>\bar{r}^{}_\perp>\bar{r}^{}_\parallel,
\end{align}
implying that $S^{n}=S^\parallel$ will be the main order parameter.
As we iterate the recursion relations (\ref{rrr}), all the $\bar{r}$'s grow with $\ell$, but $\bar{r}^{}_1(\ell)$ will reach the value 1 first, at $\ell=\ell^{}_1$. At that point the correlations in $S^\parallel$ have been eliminated, and we can use mean field theory to integrate $S^\parallel$ out of the partition function, leaving us with an effective Hamiltonian in the remaining $n-1$ spin components. We can next iterate until $\ell^{}_2$, when $\bar{r}^{}_\perp(\ell^{}_2)=1$, eliminate the $(n-2)$ transverse components with $\beta=2,\dots,n-1$ and remain with the single temperature variable $\bar{r}^{}_\parallel(\ell^{}_2)$. The scaling field for that variable finally yields
\begin{align}
T^{}_c(B)=T^{MF}_\parallel[\bar{r}^{}_\perp(\ell^{}_2)-\bar{r}^{}_\parallel(\ell^{}_2)+8u(\ell^{}_2)M(\ell^{}_2)^2].
\label{tcc}
\end{align}

The dilution is easily introduced via replicas~\cite{GrL,AAran,ma,new}. Each spin component $S^\alpha$ is replaced by $m$ replicas, $S^{\alpha i},\ i=1,\dots,m$, and at the end one takes the limit $m\rightarrow 0$. In real space, the Hamiltonian of a specific realization of the occupations, $\{p^{}_{\bf R}\}$, is
\begin{align}
{\cal H}(\{p^{}_{\bf R}\},\{{\bf S}^{}_{\bf R}\})&=-\Big[\frac{1}{2}\sum^{}_{{\bf R\ne R'}}\sum_{\alpha\beta}^n p^{}_{\bf R}p^{}_{\bf R'}W^{\alpha\beta}({\bf R}-{\bf R'})S^{\alpha}_{\bf R}S^{\beta}_{\bf R'}\nonumber\\
&+\sum^{}_{{\bf R}}p^{}_{\bf R}\big(\sum^{}_\alpha(r^{}_\alpha-r)(S^{\alpha}_{\bf R})^2-BS^{1}_{\bf R}+u|{\bf S}^{}_{\bf R}|^4\big)\Big],
\end{align}
where $W^{\alpha\beta}({\bf R}-{\bf R'})$ contains all the interactions and $p^{}_{\bf R}=1$ or $0$ with probabilities $1-x$ or $x$.
The free energy for this realization is $ F=-\log Z(\{p^{}_{\bf R}\})$, where $Z(\{p^{}_{\bf R}\})={\rm Tr}^{}_{\{{\bf S}^{}_{\bf R}\}}\exp[-\beta {\cal H}(\{p^{}_{\bf R}\},\{{\bf S}^{}_{\bf R}\})]$.

The quenched average free energy requires averaging of the free energy over the distribution of the $p^{}_{\bf R}$'s,
\begin{align}
-\beta \overline{F}=\overline{\log Z}=\overline{\lim^{}_{m\rightarrow 0}[Z^m-1]/m}=\overline{\lim^{}_{m\rightarrow 0}[Z_m-1]/m},
\end{align}
 where the overline denotes averaging over the $p^{}_{\bf R}$'s, and
 \begin{align}
 \overline{Z^{}_m}=\overline{{\rm Tr}^{}_{\{{\bf S}^{}_{\bf R}\}}\exp\Big[-\beta \sum^m_{i=1}{\cal H}(\{p^{}_{\bf R}\},\{{\bf S}^{i}_{\bf R}\})\Big]}.
 \end{align}
The exponential on the RHS contains $m$ replicas of the original Hamiltonian, each with its own spins ${\bf S}^i_{\bf R}$'s.

The average of the  free energy over the quenched random variables is achieved using the identity
~\cite{ma,AAran}
\begin{align}
\overline{e^{\bar{\cal H}}}=1+\overline{\bar{\cal H}}+\overline{\bar{\cal H}^2}/2+\dots=e^{\bar{\cal H}^{}_{ran}},
\end{align}
where
\begin{align}
\bar{\cal H}^{}_{ran}=\overline{\bar{\cal H}}+[\overline{\bar{{\cal H}^2}]^{}_{cum}/2+\dots},
\end{align}
and "cum" means the cumulant of the quenched average.
Averaging over cumulants of products of the $p^{}_{\bf R}$'s without the magnetic field we end up with
\begin{align}
\bar{\cal H}^{}_{ran}=-\sum_{i=1}^m\frac{1}{2}\int_{\bf q}U^{\alpha\beta}_2({\bf q})S^{\alpha i}_{{\bf q}}S^{\beta i}_{-{\bf q}}-\sum_{i,j=1}^m(v+u\delta^{}_{ij})\int_{\bf q}\int_{\bf q'}\int^{}_{\bf q''}S^{\alpha i}_{\bf q}S^{\alpha i}_{{,\bf q'}}S^{\beta j}_{{\bf q''}}S^{\beta j}_{{\bf -q-q'-q''}},
\label{Hr}
\end{align}
with
\begin{align}
T^{MF}_\alpha\rightarrow (1-x&)^2T^{MF}_\alpha,\ \ \ g^{}_a\rightarrow (1-x)g^{}_a,\ \ \ r^{}_d\rightarrow (1-x)^2g^{}_d,\nonumber\\
& B\rightarrow (1-x)B,\ \ \ u\rightarrow (1-x)u
\end{align}
and -- to leading order in $x$,
$v=x \Delta^2$,
where $\Delta^2$ is the mean square deviation of the `bare' transition temperature.~\cite{ma}
One can then follow Ref.~\cite{AAran}, derive recursion relations for $u$ and $v$, send $m$ to zero and follow the 3 steps that led to Eq. (\ref{tcc}).
Generally, I have shown that $T^{}_c$ is lowered by fluctuations in all three types of spins.

The term with the magnetic field adds a quadratic coupling between the replicas,
\begin{align}
\overline{\cal H}^{}_{RF}=x(1-x)B^2\sum^{}_{\bf R}\sum^{}_{i,j}S^{1i}_{\bf R}S^{1j}_{\bf R}+\dots,
\end{align}
where the dots indicate higher orders. This is the same term generated by the random field Ising model, see Sec. 2.3, but for the non-ordering spin component $S^1$. This will further lower $T^{}_c$, but the details still need detailed analysis. It should also be noted that, even without randomness, the  off-diagonal terms in ${\cal H}$, e.g. $\sum^{}_{\bf R,R'}S^{\alpha}_{\bf R}S^{\beta}_{\bf R'}$, also generate effective-field-like effects, which may  turn the problem onto the random field  model (see Sec. 2.3).~\cite{AASSC,schra,aepli} As discussed in Sec. 2.3 below, this problem is still open, and may be responsible for the loss of ferromagnetic order at low temperatures i the preset case.

As stated, quantum fluctuations can be added for low temperatures by adding an imaginary time axis. At sufficiently low temperatures, these fluctuations imply mapping of $d$ to $d+1$,~\footnote{This dimensionality shift is based on the relation $\tau\sim\xi^z$, where $\tau$ and $\xi$ are the temporal and spacial correlation lengths. Here we follow Ref. \cite{suzuki} and adopt $z=1$, as also found for the quantum one dimensional Ising model in a transverse field~\cite{elliot}. The role of fluctuations will decrease for any $z>0$, e.g., $z=2$ for $n>1$.} thus decreasing all the effects discussed above. This may explain the flattening of the curves $T^{}_c(B,x)$ at low temperatures (observed experimentally.~\cite{shechter})

So far, we only emphasized the shift in the transition temperature. After eliminating all the fluctuations, he expected critical behavior should be that of the uniaxial dipolar system in a transverse field, with (at finite temperatures) logarithmic corrections to mean field theory. The simulations in Ref. \cite{shechter} and the experiments do not yet seem to have reached this limit, so they probably observe effective exponents in the middle of the crossovers towards it.

Unfortunately, the editors of this book imposed a deadline, preventing me from finishing all these calculations. I invite everybody to do that, and -- more importantly -- extend these calculations to higher orders in $\epsilon$ or using other, more accurate methods to solve this RG project. Notably, this material seems to involve all the outstanding issues in critical phenomena, and it certainly deserves more research.

%%%%%%%%%%%%%%%%%%%%%%%%%%%%%%%%%%%%%%%%%%%%%%%%%%%%%%%%%%%%%%%%%%%%%%%%%%

\subsection{Multicritical points}

\begin{figure*}[htb]
\vspace{-1.8cm}
\includegraphics[width=1.1\textwidth]{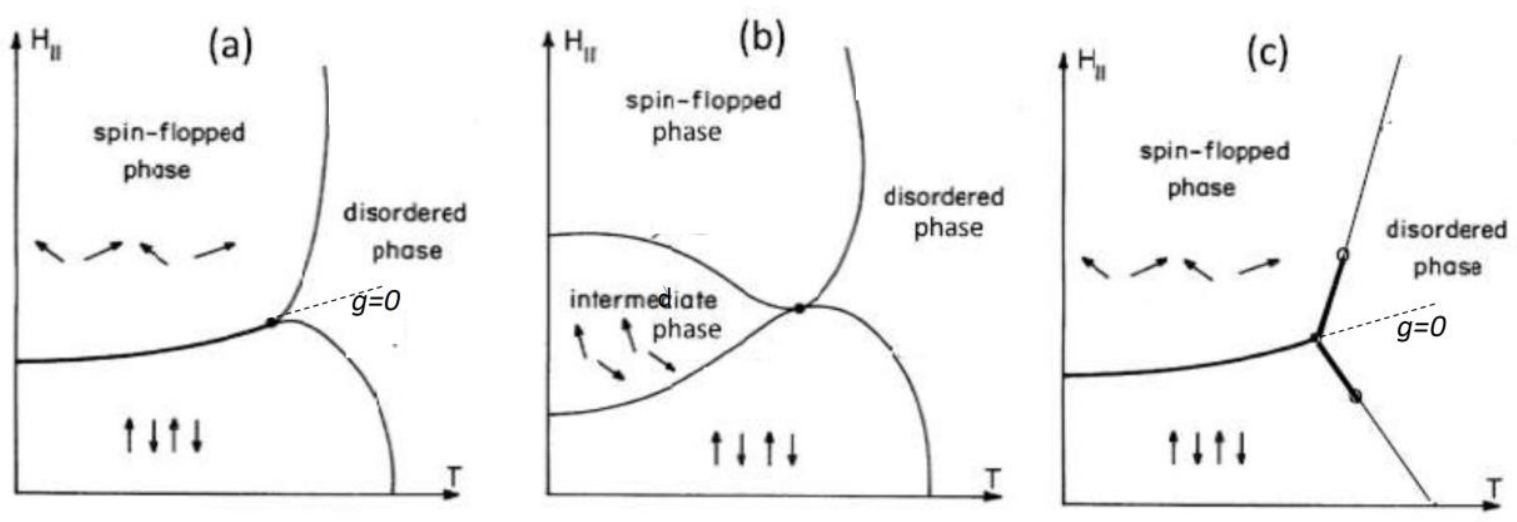}
\vspace{-15cm}
\caption{  Possible phase-diagrams for the XXZ antiferromagnet in a longitudinal magnetic field. (a) Bicritical phase diagram.  (b) Tetracritical phase diagram. (c) Diagram with a triple point.  Thick lines -- first-order transitions. Thin lines -- second-order transitions. The first-order transition lines between the ordered phases and the disordered paramagnetic phase end at tricritical points (small empty circles). After Refs. \cite{AB,DMF}. }
\label{FD}
\end{figure*}

As mentioned, the modern history of bicritical and tetracritical points, using the RG, started with Ref.~\cite{KNF}, which followed an earlier mean field paper by Liu and Fisher~\cite{liu}. The phase diagram of a system with two order parameters, having $n^{}_1$ and $n^{}_2$ components, respectively, contains two phases, in which these order parameters are (separately) non-zero. Fig. \ref{FD} exhibits three possible phase diagrams for the special case of the XXZ antiferromagnet (a Heisenberg $n=3$ antiferromagnet, with an anisotropy which prefers ordering along the $Z-$axis.)  In that example, the longitudinal magnetic field $H^{}_\parallel$ modifies the spin anisotropy, causing the transition from the longitudinally ordered phase ($n^{}_1=1$) to the `flopped' phase ($n^{}_2=2$). Experimentally and numerically, these phases are often separated by a first-order ``flop" line, which ends at a {\it bicritical} point, Fig. \ref{FD}(a). Alternatively, the two phases can be separated via second order transition lines by a ``mixed" intermediate phase, and these lines  meet the two transition lines between the ordered phases and the disordered phase at a {\it tetracritical} point, Fig. \ref{FD}(b).~\cite{KNF,AB} A third possibility, which will be the main topic of the present subsection, is shown in Fig. \ref{FD}(c): the bicritical point of Fig. \ref{FD}(a) is replaced by a {\it triple} point, at which three first order lines meet~\cite{DMF}. In this case, the transition from the disordered phase into the ``mixed" phase on the ``flop" line, along the `isotropic' line $g=0$, is also first order.

Before presenting the detailed RG arguments, here is a summary: At the multicritical point, and along the line $g=0$ in Fig. \ref{FD}, the symmetry between the $n$ spin components is maintained. For a while, it was thought that this point is described by the isotropic (rotationally invariant) Heisenberg fixed point, implying a bicritical point. However, many modern accurate techniques show that for $n=d=3$ this fixed point is slightly {\it unstable}, and the RG trajectories flow slowly either to a biconical fixed point, implying a tetracritical point, or towards a first order transition, implying a triple point. 
Neither the experiments nor the Monte Carlo simulations show either of these asymptotic predictions, hence the long-time puzzle. Below I present a novel recent way to calculate the RG trajectories, expanding around the isotropic fixed point. Since the experiments and simulations show an apparent bicritical phase diagram, we conclude that they follow the second scenario, but do not reach the triple point because of the slow flow, which requires a very large correlation length. However, close to the latter limit the effective exponents change significantly, and could be measured for checking the theory.

The nature of this multicritical point in $d=3$  has been a matter of dispute for many years. The critical properties at this point are determined by the RG analysis along the line $g=0$ in Fig. \ref{FD}. This analysis starts with a general Ginzburg-Landau-Wilson Hamiltonian (in momentum space),
\begin{align}
\overline{\cal H}=-\Big[\frac{1}{2}\int_{\bf q}\sum^{}_\alpha\big(r^{}_\alpha+q^2\big)S^\alpha_{\bf q}S^\alpha_{-{\bf q}}+\sum^{}_{\alpha\beta\gamma\delta} u^{}_{\alpha\beta\gamma\delta}\int_{\bf q}\int_{\bf q'}\int^{}_{\bf q''}S^\alpha_{\bf q}S^\beta_{\bf q'}S^\gamma_{\bf q''}S^\delta_{\bf -q-q'-q''}\Big],
\label{H1}
\end{align}
where the integrals are over a spherical Brillouin zone of radius 1. Generalizing Eq. (\ref{rra}) to the case of $n^{}_1+n^{}_2=n$ components, we have
\begin{align}
r^{}_\alpha&=r^{}_\parallel=r-(n^{}_2/n)g,\ \ \ 1\leq\alpha\leq n^{}_1;\nonumber\\
r^{}_\alpha&=r^{}_\perp=r+(n^{}_1/n)g,\ \ \ n^{}_1+1\leq\alpha\leq n.
\label{rra1}
\end{align}
The quadratic anisotropy $g$ causes the crossover from the multicritical point to the $n^{}_1-$ and $n^{}_2-$component critical lines below and above it.~\cite{pfeuty}. From now on we set $g=0$.

All the previous work started at the Gaussian FP, $u^{\ast,G}_{\alpha\beta\gamma\delta}=0$, and expanded the free energy in powers of the quartic coefficients $\{ u \}$. The recursion relation $\partial u^{}_{\alpha\beta\gamma\delta}/\partial\ell=\epsilon u^{}_{\alpha\beta\gamma\delta}-O(u^2)$ then gave non-trivial FPs, with $u=O(\epsilon)$. Depending on details, one always found four FPs: the Gaussian one, the isotropic FP, at $u^{}_{\alpha\beta\gamma\delta}\equiv u^{\ast,I(n)}=8\pi^2\epsilon/(n+8)+O(\epsilon^2)$, and the decoupled FP, with $u^{}_{\alpha\beta\gamma\delta}\equiv u^{\ast,I(n^{}_1)}$ for $\alpha,\beta,\gamma,\delta\leq n^{}_1$ and $u^{}_{\alpha\beta\gamma\delta}\equiv u^{\ast,I(n^{}_2)}$ for $\alpha,\beta,\gamma,\delta> n^{}_1$. The fourth FP depends on the details: Kostelitz, Nelson and Fisher~\cite{KNF} found the biconical FP, and in the cubic case $u^{}_{\alpha\alpha\beta\beta}\equiv u+v\delta^{}_{\alpha\beta}$ Ref.~\cite{cubic} found the cubic FP. Only one of the four FP's was stable (in the RG sense: at criticality ($T=T^{}_c$) there is a region in the $u-$space which flows under the RG iterations to that FP). The isotropic $n-$component FP was always stable (in the critical space spanned by the $u$'s) for $n$ below a threshold $n^{}_c(d)$. Based on early low order $\epsilon-$expansions~\cite{KW}, it was thought that $n^{}_c(3)>3$, and therefore the multicritical point is associated with the isotropic FP.
However, as mentioned, higher order expansions (up to $\epsilon^6$) and modern re-summations of these divergent (but Borel summable) series, as well as Monte Carlo simulations  yielded $2.85<n^{}_c(3)<3$. Therefore, the isotropic FP is unstable at $d=3$:  As $n$ crosses $n^{}_c(d)$, an operator which was irrelevant at the isotropic FP for $n<n^{}_c(d)$ becomes relevant, and the RG trajectory related to it flows away. If this flow goes to an alternative, stable FP (biconical, cubic or decoupled), the transition at the multicritical point is continuous, and is then identified as a tetracritical point (at these points the minimum of the quartic Hamiltonian below the multicritical point has a mixed state~\cite{liu}).
Below we return to what happens when the flow does not reach a stable FP.

Until now, no experiments nor simulations observed any sign of a first order transition at the multicritical point. If indeed the transition is second order,
it must be associated with one of the above stable FP's, hence the tetracritical scenario. This conclusion
generated  Ref.~\cite{mef70}, commenting on a Monte Carlo study~\cite{hu} of Zhang's $SO(5)$ model~\cite{zhang} for the high temperature superconductors, which included the $n^{}_1=3-$component antiferromagnet and the $n^{}_2=2-$component superconductor. In that case there exists an exact proof that the stable FP is the decoupled FP, and my comment wondered why the $SO(5)$ did not find that, and why they seemed to observe a bicritical point with exponents associated with the isotropic FP.
 Obviously, the late Shoucheng Zhang was not happy with that comment. However, I did end the comment saying: ``All the above statements assume that the initial Hamiltonian is in the region of attraction of the decoupled FP. Alternatively, one should expect a first order transition... The apparent experimental observation of isotropic exponents may still indicate that the initial Hamiltonian is close to the isotropic FP." %So far, nobody observed a first order transition, and the observed exponents do seem close to those of the isotropic FP, apparently in agreement with those sentences. However, a  reliable calculation was still needed to confirm this and to connect the vicinity of the isotropic FP with the possible first order transition.
%Sice that nobody at that time found a first order transition, left an open puzzle
Indeed, the calculation below confirms these last comments: if one has a finite sample of linear size $L$ then after a finite number of RG iterations $\ell=\ln L$ the RG flow stays near the isotropic FP, and does not reach the triple point.

In many earlier papers,
 when the RG trajectories flowed away from an unstable point, without reaching any alternative stable FP, it has been generally expected that  the renormalized effective free energy, which can be solved after all the fluctuations have been removed, has a first order transition. If the transition at the multicritical point (along the line $g=0$ in Fig. \ref{FD}) becomes first order, then (by continuity) the two transition lines from the disordered phase also become first order in its vicinity, and the multicritical point then becomes a {\it triple point}~\cite{DMF}. These first order regions then end at {\it tricritical} points, Fig. \ref{FD}(c).  However, the experiments and simulations have not yet exhibited such triple points. Also, there did not exist accurate estimates on how close to criticality should the triple point arise.  Hence the ``puzzle of the bicritical points", which I studied recently with my long-time colleague Ora Entin-Wohlman~\cite{AE}, based on calculations with Andrey Kudlis~\cite{AEK1,AEK2,AEK3}, who was a major player in the re-summed high order $\epsilon-$expansions~\cite{6eps}. Since these references give full details of our calculations, I give here only a brief summary.

 Generally, the recursion relation near a FP for a parameter $G^{}_i$ has the form
 \begin{align}
\partial g^{}_i/\partial\ell=\overline{\beta}^{}_i[g^{}_1,~g^{}_2,\dots].
\end{align}
In our case, the $\overline{\beta}-$functions are generated as sixth order expansions in the $u's$, and people solved them for the FP's. However, we need more that just the FP's. To follow the RG trajectories in the $u-$space we need the actual variation of $\overline{\beta}^{}_i$ on its $\ell-$dependent arguments, and (like the $\epsilon-$expansion) the series in these arguments do not converge! For $n=3$ we know that the isotropic and biconical (or cubic) FP's are close to each other, and therefore we decided to construct recursion relations in the vicinity of the isotropic FP, rather than near the Gaussian FP. Thus, we expanded $\overline{\beta}^{}_i$ in powers of $\delta g^{}_i=g^{}_i-g^{\ast,I(3)}_i$ around the isotropic FP. The coefficients in these expansions are derivatives of $\overline{\beta}^{}_i$ with respect to the various $g^{}_j$'s, at the isotropic FP, which are known series in the $g^{\ast,I(3)}_i$. The latter series can be written as series in $\epsilon$, and can then be resummed to give good estimates of their accurate values at $d=3$. Close to the isotropic FP it was sufficient to use only second order expansions in the deviations $\delta g^{}_i$'s; the resulting values of the cubic or of the biconical FP's were quite accurate.

We now turn to our main task: what happens if we start with a Hamiltonian out of the region of attraction of the stable (biconical or cubic) FP?
Group theory~\cite{wegner,zan,vicari,MC} shows that there exist only three independent stability exponents near the isotropic $n-$component FP. The quartic terms in Eq. (\ref{H1}) split into three subgroups of the rotationally invariant $O(n)$ group. For even functions in the spins, the quartic terms can always be written as
\begin{align}
{\cal H}^{}_4=(u^{\ast,I(n)}+p^{}_0){\cal P}^{}_0+p^{}_2{\cal P}^{}_2+p^{}_4{\cal P}^{}_4,
\end{align}
where the $p^{}_i$'s are scaling fields which flow under the RG iterations, with (for $d=3$) stability exponents
(agreed by all the extrapolations)
 ~\cite{boot}
 \begin{align}
 \lambda^{}_0\approx -0.78,\ \ \lambda^{}_2\approx -0.55,\ \ \lambda^{}_4\approx 0.01.
 \label{exps}
 \end{align}
The operator ${\cal P}^{}_0\equiv|{\bf S}|^4$ is the single fully isotropic quartic term, which characterizes the flow to the isotropic FP. The other operators break rotational symmetry.
The operator ${\cal P}^{}_2$ is a linear combination of operators which `prefer' ordering of one of the competing order parameters, e.g.,
${\cal P}^{}_2=|{\bf S}|^2\big(n^{}_2|{\bf S}^{}_\parallel|^2-n^{}_1|{\bf S}^{}_\perp|^2\big)$.
For $n=3$ there are five such operators. As we said, the multicritical point requires `isotropy', hence $g=p^{}_2=0$. The remaining 9 operators form ${\cal P}^{}_4$, which (with no cubic terms, $v=0$) has the form
\begin{align}
&{\cal P}^{}_{4}\equiv\mid{\bf S}\mid^4\Big[\frac{n^{}_1n^{}_2}{(n+2)(n+4)}
+x(1-x)
-\frac{n^{}_1(1-x)+n^{}_2 x}{n+4}\Big],\ \ \ x=\frac{|{\bf S}^{}_{\parallel}|^2}{|{\bf S}|^2}.
\label{P4}
\end{align}
It is easy to express the original $u$'s in terms of the three $p$'s, and thus find the recursion relations for the $p$'s.

Since we set $p^{}_2=0$, we are left with the coupled recursion relations for $p^{}_0$ and $p^{}_4$. Since $|\lambda^{}_0|$ is large, $p^{}_0(\ell)=p^{}_4(0)e^{\lambda^{}_0\ell}$ decays fast, and can be neglected after $\ell^{}_1$ iterations.
Beyond that point, the quadratic recursion relation for $p^{}_4$ can be solved analytically, yielding~\cite{AEK1,AEK3}
\begin{align}
p^{}_4(\ell)=\frac{p^{}_4(\ell^{}_1)e^{\lambda^{}_4(\ell-\ell^{}_1)}}{1+B p^{}_4(\ell^{}_1)\big[e^{\lambda^{}_4(\ell-\ell^{}_1)}-1\big]/\lambda^{}_4},
\label{vvv}
\end{align}
where the biconical (or cubic) FP is at $p^{\ast,B}=\lambda^{}_4/B>0$. Since $\lambda^{}_4>0$ is small, the initial flow is very slow, and the trajectory stays
near the isotropic FP for many iterations. This may explain why Hu's Monte Carlo simulations of Zhang's $O(5)$ model gave an apparent  bicritical diagram with isotropic  exponents. For $p^{}_4(0)>0$, the trajectory goes to the biconical (or cubic) FP, but then one should immediately see tetracritical behavior.

The situation changes for $p^{}_4(0)<0$. In that case the trajectory again remains close to the isotropic FP, but at some point the denominator in (\ref{vvv}) vanishes, and $|p^{}_4|$ diverges. Although our quadratic approximation in the recursion relation may no longer be valid, it is clear that $p^{}_4$ varies strongly before this divergence. In Refs.~\cite{AEK1,AEK2,AE,AEK3} we show that if the iterations stop at $\ell^{}_2$, when $r(\ell)\sim 1$, then we can use mean field theory and identify when the transition becomes first order, ending us with the triple point of Fig. \ref{FD}(c).

\begin{figure}[t]
\includegraphics[width=0.45\linewidth]{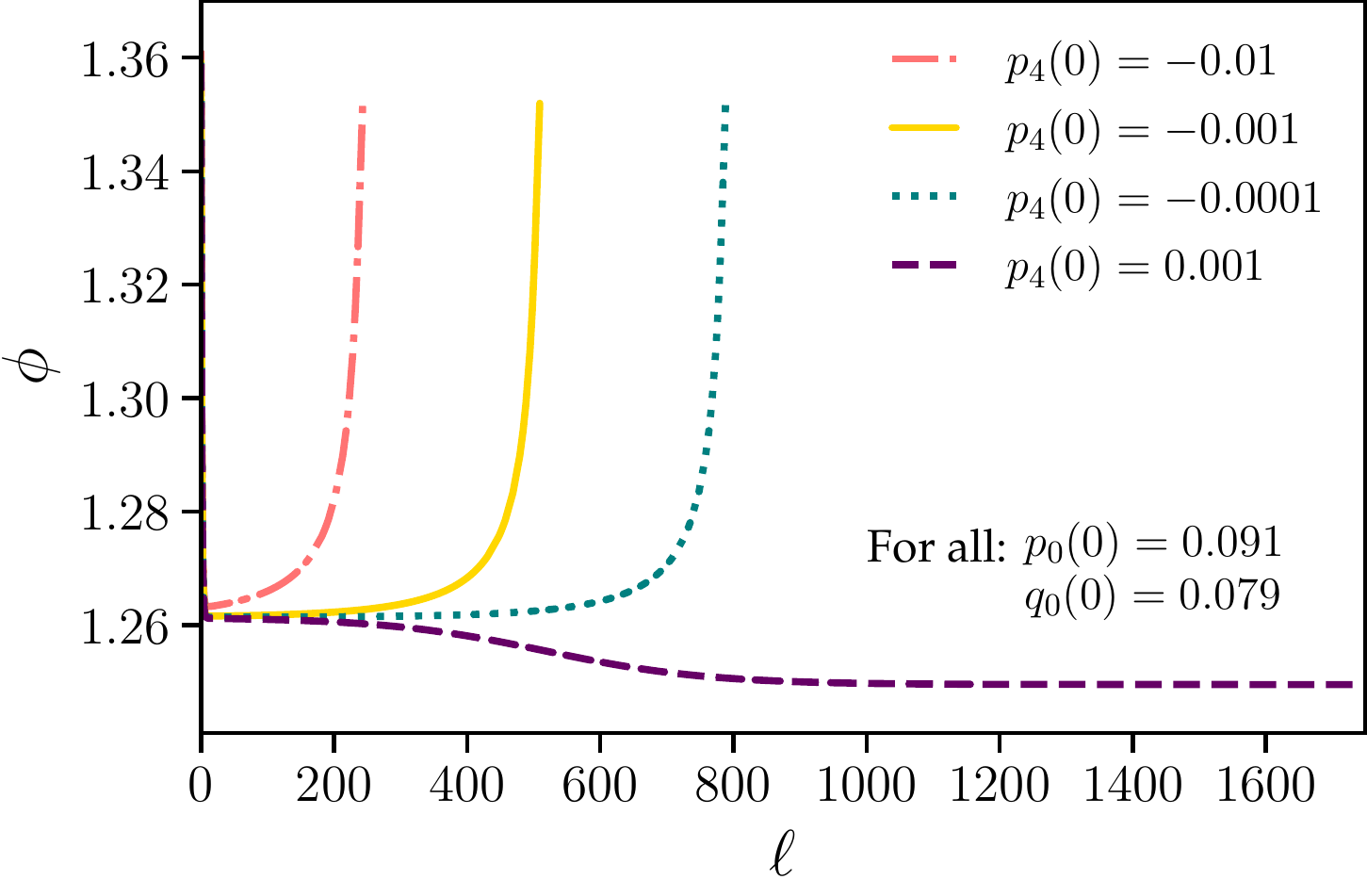}\ \ \ \ \ \ \includegraphics[width=0.45\linewidth]{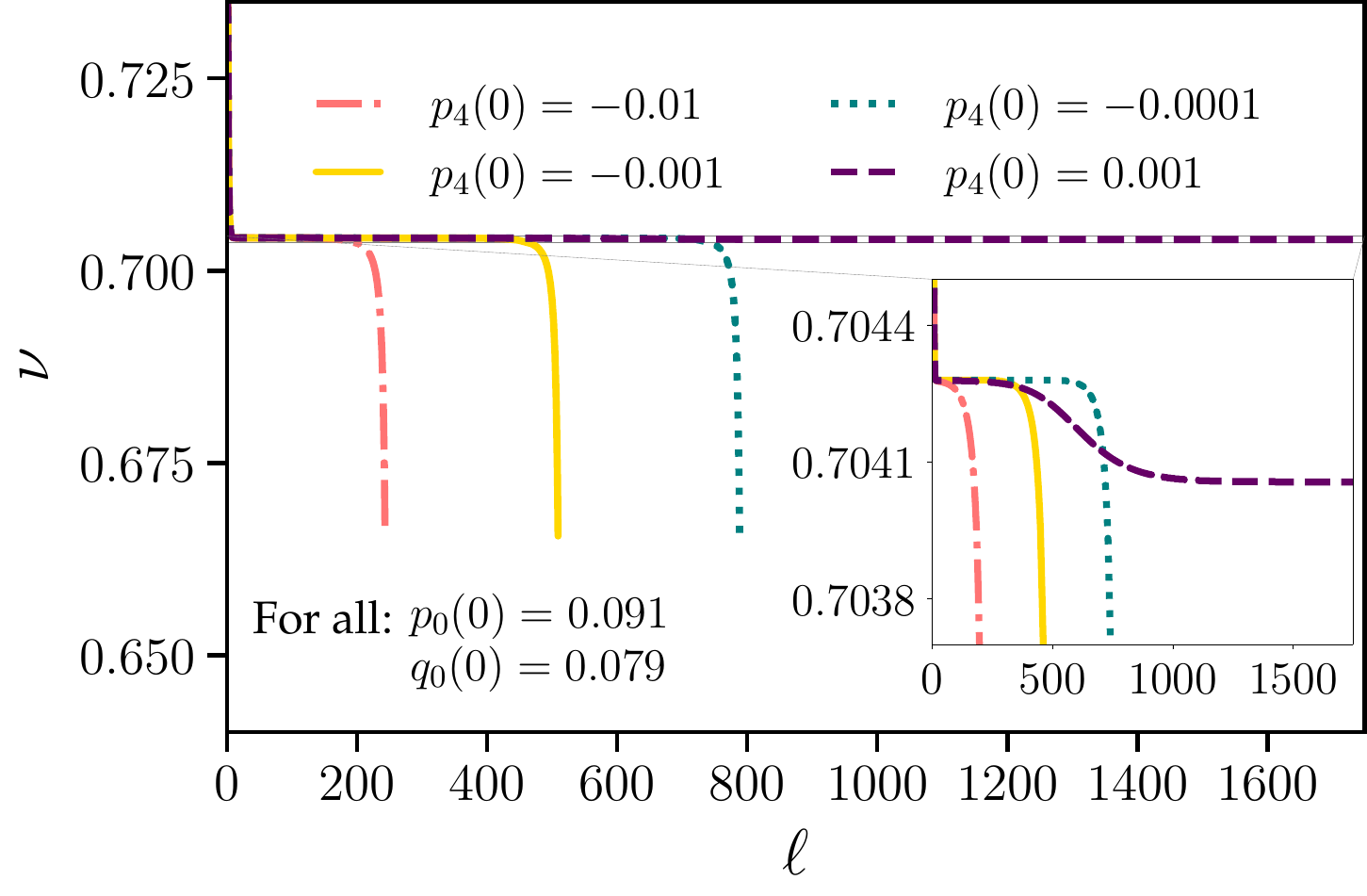}
\caption{%(color online)
 Dependence of the effective critical exponents $\nu$ and $\phi$ on the RG flow parameter $\ell$. Different lines correspond to different initial values  $p_4(0)$, for a fixed $p_0(0)$.
  In this plot, the trajectories are shown only at $p_4(\ell)>-.8$, where our quadratic approximation is reasonable. From Ref.~\cite{AEK3}.}
\label{exp}
\end{figure}

Once we have the $\ell-$dependent $p$'s, we can also substitute them into other recursion relations, e.g. for $r$ and $g$, and derive the exponents for the associated scaling fields, $t(\ell)=e^{\ell/\nu(\ell)}t(0)$ and $g(\ell)=e^{\ell\phi(\ell)/\nu(\ell)}g(0)$. Figure \ref{exp} shows these effective exponents, for the case $n=1+2$ without cubic symmetry, and demonstrates their fast variation as the bicritical point approaches the triple point.

All these calculations should be extended to higher orders in the $\delta g^{}_i$'s. In our papers we also suggest how one may change the parameters, in order to approach the triple point and demonstrate Fig. \ref{FD}(c) by experiments or simulations.

%%%%%%%%%%%%%%%%%%%%%%%%%%%%%%%%%%%%%%%%%%%%%%%%%%%%%%%%%%%%%%%%%%%%%%%%%%

\subsection{Random fields Ising model (RFIM)}

Quenched random fields exist in many physical systems~\cite{FA,cardy,deG}. Experiments are difficult, due to equilibration problems~\cite{rohrer,expts,cowley}, but still very desired. Note that some of these experiments study the drastic effects of the random fields on bicritical phase diagrams (e.g., destroying XY order below the random field lower critical dimension $d^{}_l=4$).  This model was first solved for the $n\rightarrow\infty$ limit (random sources and sinks of superfluid particles) by Lacour-Gayet and Toulouse~\cite{lacour}, who found the new upper and lower critical dimensions $d^{}_u=6,~d^{}_l=4$. Imry and Ma~\cite{IM} then gave qualitative arguments that showed that below a lower critical dimension $d^{}_l$, random magnetic fields cause a breakdown of the system into domains with opposite spins, hence a destruction of the magnetic long range order. The size $L$ of the domains is determined by the competition between the gain in bulk energy ($\pm h^{}_0 L^{d/2}$, $h^{}_0$ is the typical random field) and the loss in boundary energy  ($JL^{d-1}$ for $n=1$ or $JL^{d-2}$ for $n>1$; $J$ is the exchange energy). They found $d^{}_l=4$ for the rotationally invariant models (with $n\geq 2$) and $d^{}_l=2$  for the Ising case ($n=1$). In both cases, the upper critical dimension was found to be $d^{}_u=6$. As experimental realizations, Imry and Ma suggested antiferromagnetic magnetic impurities in ferromagnets, or lattice distortions.  Following Ref. \cite{FA}, many experiments were done with uniform magnetic fields on dilute antiferromagnets.

At dimensions $d>d^{}_l=2$, there exists an ordered ferromagnetic phase below $T^{}_c(B)$ for the dilute antiferromagnet in a field $B$, but the Imry-Ma domains still exist above the transition, with typical size $L\sim (h^{}_0/J)^{2/(d-2)}$, representing metastable states which take a long time to equilibrate (via avalanches of flopping domains) for large $h^{}_0$. This typical size decreases with $h^{}_0$, eventually allowing long rage order. Indeed, Ref. \cite{nowak} identified a temperature $T^{}_i(B)>T^{}_c(B)$, below which these domains cause slow dynamics. This may explain the difficulties of experiments to equilibrate~\cite{rjb}. As far as I know, there is not yet a complete consistent theory to describe this scenario.

Looking at the diagrammatic expansions in the quartic spin terms and in the random fields, it was shown that each of the leading diagrams in the random fields model in $d$ dimensions maps exactly onto a similar diagram for the non-random model in $d-2$ dimensions~\cite{AIM,young}. It was thus concluded that to all orders, the $\epsilon-$expansion of the random field model in $\epsilon=6-d$ is the same as for the non-random model with $\epsilon=4-d$~\cite{AIM}. This dimensionality shift by 2 was later found to be an exact result of supersymmetry~\cite{parisi}. Indeed, this dimensionality shift is consistent with the Imry-Ma argument for  $n\geq 2$, where the lower critical dimension without the random fields is equal to $2$.

However,  the rule $d\rightarrow d-2$ does not agree with the Ising case, where Imry and Ma predicted $d^{}_l=2\neq 1+2$ ($1$ is the lower critical dimension for the Ising model without the random fields). Also,  Imbrie~\cite{imbrie} and Brickmont and  Kupiainen~\cite{brickmont}  showed that -- unlike the non-random Ising model at $d=1$ -- there is  long range order for the RFIM at  $d=3~(=1+2)$. This disagreement with dimensionality shift by 2 has been a source of much research in the last 45 years, suggesting a variety of explanations. Some of these are reviewed below, but may questions still remain open for future research. %The non-expert reader can skip the following details, and `jump' to the end, where I list some open issues. 

In addition to the problem with the dimensionality shift, there were two other (related) questions: (1) Do all the random field distributions belong to the same universality class? and (2) Is the phase transition of the RFIM always continuous?
The literature studied several distributions of the site random fields, e.g., the Gaussian, $P(h^{}_i)\sim\exp[-(h^{}_i/h^{}_0)^2/2]$ and the bimodal, $P(h^{}_i)\sim [\delta(h^{}_i+h^{}_0)+\delta(h^{}_i-h^{}_0)]/2$.
Mean field theory~\cite{AARF,andelman} showed that the transition can be continuous at all temperatures for a distribution with a maximum at $h^{}_i=0$, like the Gaussian one, but it must become  of first order at low temperatures (and high random fields) for distributions with a minimum at $h^{}_i=0$, like the bimodal one.

To test the issue of the dimensionality shift for the RFIM, in 1993 we returned to the old and good method of high temperature series~\cite{shapirs}. As reviewed in this book by Singh, Michael Fisher played a crucial role in the development of this method (see also his list of scientific achievements at the end of this book). He is also acknowledged in this paper.\footnote{The other acknowledgement was to Brooks Harris, with whom I have been fortunate to publish many papers, both on series and on the RG.} Series for the susceptibility were developed for $d=2,3,..,8$. The results confirmed mean field exponents for $d\geq 6$ (logarithmic corrections could not be seen), and it was difficult to observe  transitions at $d=2$ and $d=3$. It was emphasized that when the random fields are switched on, one has a crossover from the non-random to the RFIM behavior (without random exchange, the crossover exponent is equal to the non-random susceptibility exponent $\gamma$~\cite{shapir2}. Random exchange causes small corrections). At large random fields and low temperatures the analysis is erratic, probably hinting at a  first order transition, and this leaves only a intermediate range of fields for extracting the RFIM exponents. At $d=5$ the result was $\gamma\approx 1.18$, but it was difficult to give error bars. This value is smaller than the $d=3=5-2$ non-random $\gamma(d=3)\approx 1.241$, but the uncertainties led the authors to say that dimensionality shift may hold at $d=5$. Results were not clear at $d=3$.

To clarify this we produced much better  series  in 1996~\cite{gofman}.
After stating that ``series expansions have a great advantage over Monte Carlo simulations, since (like the real experiments) they do not suffer from metastability and equilibration problems" [see Michael Fisher's comments on Monte Carlo simulations in Domany's paper, this book~\footnote{However, Fisher withdrew that ``slip of the tongue" in 1999 \cite{MEF1999}.}], Ref.~\cite{gofman} presented new high temperature series for the RFIM in dimensions $d=3,4,5,8$. Results at $d=8$ gave mean-field exponents, as expected for $d^{}_u=6$. Results at $d<6$ confirmed the two-exponent (and not three exponent, see below) scenario, but the numerical values of the exponents and their error bars did not seem consistent with the $d\rightarrow d-2$ rule. For example, in $d=4,~5$ we found the susceptibility exponent for {\it both} field distributions in the ranges  $\gamma=1.45\pm .05,~1.13\pm .02$, distinct from the non-random values $\gamma=1.75,~\sim 1.241$ for the non-random Ising model at $d=2,~3$. In addition, the range of temperature for which the series were stable was shorter for the bimodal distribution, which may have hinted at an approach to a tricritical point and a first order transition at low temperatures (presumably including $T=0$). We conclude that dimensionality shift indeed breaks down somewhere near $d\sim 5$, and that the critical exponents seem to come from another, supersymmetry breaking,  fixed point. 

\begin{figure}[t]
\vspace{-2.1cm}
\includegraphics[width={13cm}]{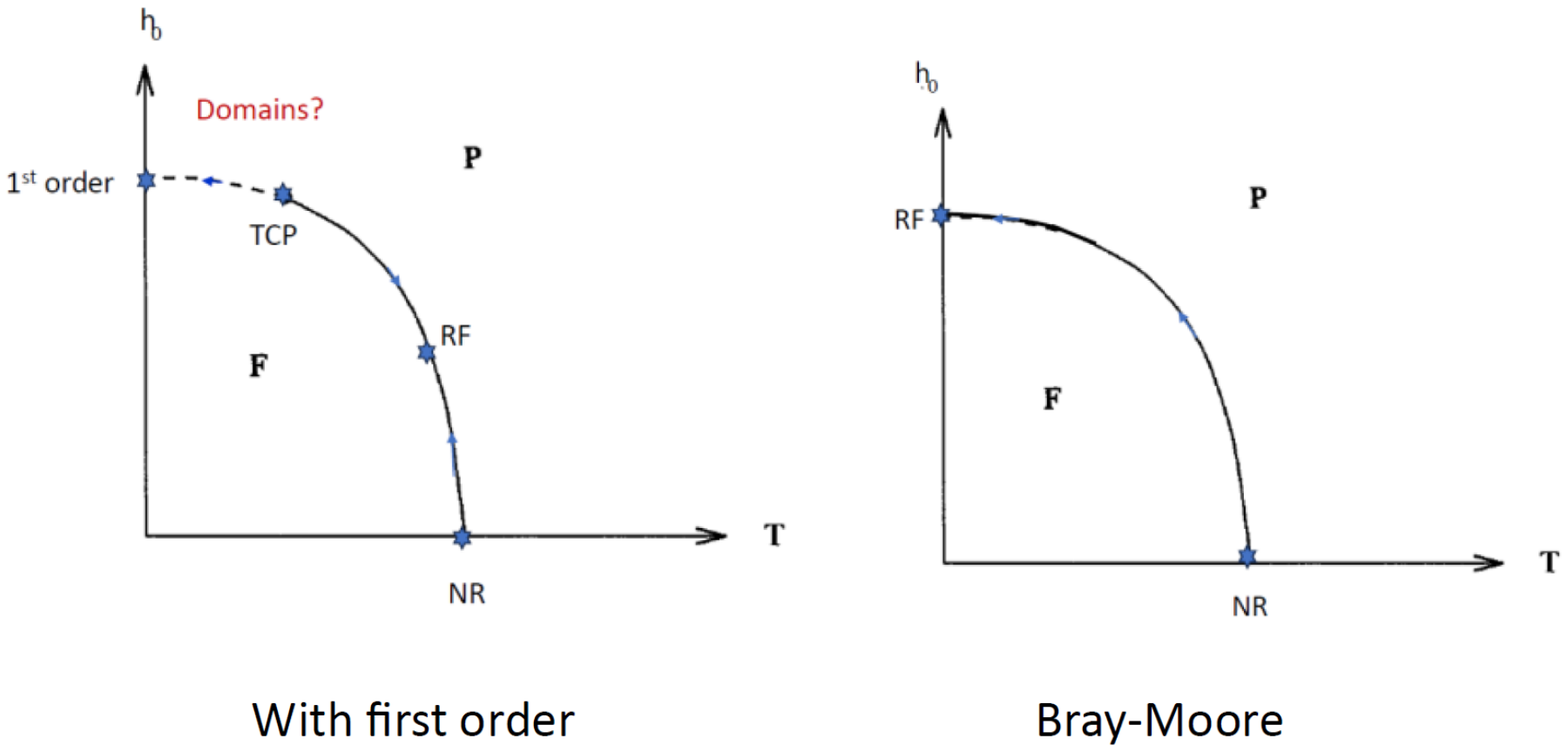}
\vspace{-11.5cm}
\caption{%(color online)
 Schematic phase diagrams in the temperature-random field $T-h^{}_0$ plane for $d=3$. Right: The random field fixed point is at zero temperature. Left: conjectured diagram for the binary and similar field distribution. P=paramagnetic (but with Imry-Ma domains), F=ferromagnetic (or any other ordered Ising phase).  NR=non-random,  RF=random field, TCP=tricritical point. $h^2_0$ is the mean square average of the field distribution. The blue arrows indicate the renormalization flow at criticality. }
\label{PDAA}
\end{figure}

Many theories tried to generalize the $d\rightarrow d-2$ rule in the hyperscaling relations (e.g., $d\nu=2-\alpha$) by a new rule, $d \rightarrow d-\theta$, breaking supersymmetry and the dimensionality shift. This procedure  introduced a new  independent exponent $\theta$ (presumably related to the dimensionality of the roughened domain boundaries~\cite{grinstein}). These models thus require three (instead of the usual two) independent critical exponents. Some of these models, which involve zero temperature fixed points,  were reviewed e.g. in  Ref.~\cite{gofman}. The non-random fixed point associated with the transition as function of $T$ at $T^{}_c(0)$, is strongly unstable against the random fields, and the RG trajectories flow from that fixed point to the RFIM fixed point, associated with the transition as function of $h^{}_0$ at $T=0$, see RHS of Fig. \ref{PDAA}. Bray and Moore~\cite{bray} use simple arguments to identify the recursion relations near that fixed point, and argue that it is stable with respect to the temperature. Therefore, thermal fluctuations are irrelevant, and it is sufficient to find the ground state as function of the random fields at $T=0$. Such a scenario would replace the supersymmetric fixed point at low dimensions. However, it remained to find how the supersymmetric fixed point turns into this new fixed point. Also, this scenario, which assumes a continuous transition,  cannot be valid for the binary field distributions, which should yield a first order transition (as predicted by mean field theory; fluctuations may narrow the range of such a transition, but not eliminate it completely). As explained in \cite{AARF}, for the binary distributions there must exist an ustable tricritical point on the critical line of the RFIM, between the non-random fixed point at $T^{}_c(h=0)$ and he one at $T=0$. This would require a stable RFIM fixed point at a finite temperature between the non-random fixed point and the tricritical fixed point, as plotted on the LHS of Fig. \ref{PDAA}. This phase diagram differs significantly from that of Bray and Moore. at temperatures below the tricritical fixed point, the RG trajectories would have to flow to a zero-temperature fixed point which represents a first order transition~\cite{1st-order}, which is presumably different from the Bray-Moore RFIM fixed point.
 The apparent universality found in both series and Monte Carlo must imply that all distributions are described by this finite$-T$ RFIM fixed point, and not by the zero temperature fixed point. Although some candidates for this fixed point have been proposed (see below), the full details of such an RG scenario have not yet been investigated.

 Some of the above questions were already addressed in 1997 by Swift {\it et al.}~\cite{swift}.  They used a finite size scaling analysis of the exact ground states of small RFIM samples in three and four dimensions, and found a difference between the Gaussian and the bimodal field distributions in four dimensions, with the latter apparently having a discontinuous jump in the magnetization. The two distributions generated very different distributions of the magnetization. Although their analysis was restricted to zero temperature, they did not rule out a possible RFIM fixed point at a finite temperature.
A similar study at $T=0$ by Sourlas~\cite{sourlas}, at $d=3$, indicated that the Gaussian and the bimodal distributions do not belong to the same universality class. He noticed important differences between individual realizations (which showed discontinuities)  and their averages, but still concluded that all the distributions give continuous transitions.

The last point raises the important issue of self-averaging. Away from criticality, the central limit theorem predicts that the width of the distribution of ay measured quantity decays with system size $L$ as $(\xi/L)^d$, where $\xi$ is the correlation length. However, near a random (exchange or field)  fixed point this width approaches a constant~\cite{self}. In particular, for the RFIM fixed point near six dimensions this width is proportional to $(h^2_0 u)^\ast$ at the random field fixed point, and independent of $L$, where $h^2_0$ is the second cumulant of the field distribution and $u$ is the coefficient of the non-random quadratic spin term~\cite{self}.
Since for large $L$ this width becomes finite and independent of $L$, simulations must be very careful with averages over small numbers of realizations. Except for a few examples (below), I am not aware of systematic studies of self-averaging for the general RFIM problem. Such studies might explain the different results found by different groups.

One exception is the 2002 work of Parisi and Sourlas~\cite{ParSour}. They studied the correlation functions of the three dimensional RFIM, and found that the correlation length (at finite $L$) is not self-averaging.  They blamed this result on the nonperturbative formation of a bound state in the underlying replicated field theory. This nonperturbative result may replace the supersymmetric fixed point at low dimensions, but it is not clear what are the predictions for the critical behavior.

Assuming the Bray-Moore scenario, of a continuous transition at all $T$, Ref. \cite{RFMC} performed Monte Carlo simulations at zero temperature. Apparently, such simulations avoid some of the difficulties in previous simulations. At $d>5$ these simulations agreed with the supersymmetry predictions. However, for $d<5$ they found deviations from supersymmetry.
Like our series results, these authors also found the same exponents from both distributions. However, they found the zero temperature transition continuous for all the distributions -- in contrast to the mean field theory for the bimodal distributions. As stated, such distributions require more research.
A possible way out may come from the effects of random quenched fluctuations on first order transitions. Indeed, Imry and Wortis~\cite{IW} presented a domain argument similar to that of Imry and Ma, to show that randomness (e.g., random interactions or random fields) can turn a first order transition into a continuous one (the system breaks into domains of the two competing phases). This argument was supported by various other arguments~\cite{Aizen,Berker}. However, all of these arguments still required some finite temperature below which the argument fails, and there should still be a first order transition at low temperature. Hopefully, the above story will stimulate  theoreticians (and experimentalists!) to look into this.
These authors also checked self-averaging~\cite{nikos}, and found that the widths of the susceptibility distribution approaches  finite $L-$idependent values, as predicted in Ref.~\cite{self}. However, they find different values for different distributions, which may indicate different universality classes.

The absence of self-averaging implied that the average description may differ significantly from the behavior of a single realization. Interestingly, Kumar {\it et al.}~\cite{kumar} simulated the zero temperature configurations of RFIM with a uniform ordering magnetic field at $d=3$, for specific field realizations. The domain shapes seemed to depend on the field distributions. However, they measured the dependence of the magnetization $M$ on the uniform field $H$, and identified the universal (distribution-independent) critical exponents $\beta$ and $\delta$ [$M\sim(T-T^{}_c)^\beta$ at $H=0$, $M\sim H^{1/\delta}$ at $T=T^{}_c$], finding $\beta=.025\pm .005$ and $\delta=72\pm 4$. Remembering that a first order transition would have $\beta=0,~\delta=\infty$, I wonder whether this allows for the phase diagram on the LHS of Fig. \ref{PDAA}, alas for specific realizations~\cite{specific}. It is not clear yet how to relate such realization with the averages used in all the theories.

Apart from the above numerics, there have also been  some analytic calculations, within the field theory ad the Wilsoin-Fisher renormalization group approach.
In 1998, Br\'{e}zin and De Dominicis~\cite{br-deDom} noted that the breakdown of supersymmetry should arise due to additional operators, which break supersymmetry,  to the Parisi-Sourlas Lagrangian~\cite{parisi11}. If the supersymmetric fixed point becomes unstable with respect to such operators, this would explain the breaking of supersymmetry. However, they found a singular dependence on the replica number $n$, ending up with a runaway from the supersymmetric fixed point, without any other, stable,  fixed point.

In 2004, Tarjus and Tissier~\cite{TT} started a long series of papers, in which they reformulated the functional nonperturbative RG in a superfield formalism, generalizing the Parisi-Sourlas supersymmetric theory.
They concluded that the failure of dimensionality reduction and standard perturbation theory below $d\simeq 5.1$ is due to the nonanalytic nature of the zero-temperature `cuspy' fixed point, which is associated with the existence of many metastable states and the avalanches between them. The cusps appear in the cumulants of the renormalized field distributions. Near the zero temperature fixed point, the temperature is a dangerous irrelevant field, related to the appearace of droplets, which cause rounding of the cusps.

%Another exciting result came from recent  field-theoretical considerations, combining nonperturbative conformal field theory with perturbative RG~\cite{RFslava}. These calculations indicated that the dimensionality shifted FP, which is clearly stable in $6-\epsilon$ dimensions, becomes unstable, and replaced by another, stable, FP, below $d\sim 4.5$. As these authors say, this situation is similar to that discussed in the previous subsection, where the $n=3$ isotropic FP becomes stable at a dimension slightly below 3, replacing the biconical (or cubic) FP. The new FP arises from non-supersymmetric operators, which were irrelevant at the supersymmetric FP in $d>4.5$, but became relevant at lower dimensions. How this new FP develops at lower dimensions, and does it disappear at the  Imry-Ma $d^{}_l=2$, remains an open question.

Another exciting result came from a recent theoretical study, which combined nonperturbative conformal field theory with perturbative RG~\cite{RFslava}. These authors also added interactions which break supersymmetry, but are irrelevant near the supersymmetic fixed point near $d=6$. Examples include quartic spin terms, which may become relevat below $d=4$. They found that two such operators become relevant below $d^{}_c\approx 4.2-4.7$, indicating a crossover to a new, stable, fixed point, at which supersymmetry is broken.      As these authors say, this situation is similar to that discussed in
the previous subsection, where the $n = 3$ isotropic fixed point becomes stable at
a dimension slightly below 3, replacing the biconical (or cubic) fixed point (which are stable at higher dimensions). Since this calculation uses the expansion near $d=6$ down to the vicinity of $d=4$, it still requires support from other approximations. In any case, how this
new fixed point develops at lower dimensions, how it relates to the phase diagrams in Fig. \ref{PDAA}, and does it disappear at the Imry-Ma
$d^{}_l = 2$, remain  open questions.

In summary, here are some of the remaining open questions:

(a) Which of all the suggested fixed points replaces the supersymmetric fixed point, and breaks supersymmetry? Are essential singularities important in this context?

(b) Is that fixed point at zero or at a finite temperature? 

(c) If the LHS of Fig. \ref{PDAA} is correct, what are the details of the tricritical point and of the first-order fixed point?

(d) How does that fixed point relate to the Imry-Ma scenario?

(e) Is that fixed point universal, for all field distributions?

(f) Does that fixed point confirm the breakdown of self-averaging? If so, how does it relate to observations for single field realizations?

(g) How can the slow dynamics and avalanches of domain flipping be incorporated into the renormalization group treatment? 

(h) All the above was for symmetric field distributions, $P(h)=P(-h)$. There were claims that non-symmetric distributions, e.g., for a binary mixture in a porous media,  may yield a different behavior,~\cite{maritan} but this issue deserves further research.

Surely, more work will discover may more questions. There is much more to do on the RFIM.

\section{Conclusions}

I hope that my story has convinced you that Michael E. Fisher was a great scientist, a great leader of the statistical physics community, and a great mentor and teacher. I owe him a lot.

I also hope that I convinced you that both the life of Michael Fisher and the golden jubilee of the $\epsilon-$expansion deserve a celebration.
Although the life of Fisher ended, the $\epsilon-$expansion and other RG methods are still very active, as I hope I demonstrated by my few examples. There is much more to do!

\vspace{1cm}%%%%%%%%%%%%%%%%%%%%%%%%%%%%%%%%%%%%%%%%%%%%%%%%%%%%%%%%%%%%%%%%%%%%%%%%%%
\noindent{\it \bf Acknowledgements}:

\vspace{0.7cm}

I thank Ora Entin-Wohlman and Andrey Kudlis for collaboration, and Slava Rychkov and Nikos  Fytas for very useful discussions.
Eytan Domany and David Nelson kindly helped checking the history and identifying people in Fig. 2, Jacques Perk caught many misprints and added some references, and Leo Radzihovsky corrected some crucial errors in an earlier version. I apologize to the many people whose references were not included, only due to space and time constraints.

%\newpage
%\section*{Appendix}


\begin{thebibliography}{99}

\bibitem{60} {\bf Current problems in statistical mechanics}, ed. by E. Domany and D. Jasnow,  Physica A {\bf 177}, Issues 1-3 (1991).

\bibitem{70} J. Stat. Phys. {\bf 110}, Special issue in honor of Michael E. Fisher's 70th birthday, (edited by Joel Lebowitz), Numbers 3-6 (March 2003).

\bibitem{80} 106th Rutgers Statistical Physics Conference, Dec. 2011: https://cmsr.rutgers.edu/docman-lister/cmsr/events-cmsr/
    statistical-mechanics-conference/106/1070-smm106-finalprogram/file


\bibitem{stellen} {\bf Critical Phenomena, Proc. Stellenbosch, S. A. 1982}, F. J. W. Hahne, ed., (Springer, Berlin, 1983).

\bibitem{MEFq} Michael E. Fisher, {\it Condensed Matter Physics: Does Quantum Mechanics Matter?}, Excursions in the Land of Statistical Physics, pp. 207-253 (World Scientific 2016). Reprinted  from “Niels Bohr: Physics and the World,” Eds. H. Feshbach, T. Matsui and A. Oleson (Harwood Academic Publisher, Chur, 1988) pp. 65–115.

\bibitem{WF} Kenneth G. Wilson and Michael E. Fisher, {\it Critical Exponents in 3.99 Dimensions},
Phys. Rev. Lett. {\bf 28}, 240 (1972).

\bibitem{MEFRMP} Michael E. Fisher, {\it Renormalization group theory: Its basis and formulation in statistical physics}, Rev. Mod. Phys. {\bf 70}, 653 (1998). See Also, Michael E Fisher, {\it Renormalization group theory, the epsilon expansion and Ken Wilson as I knew him},  Int. Jn Mod. Phys. B {\bf 29}, 1530006 (2015). Selected reviews by Fisher appeared in {\bf Excursions in the land of statistical physics} (World Scientific, 2017); see also Cyril Domb's paper there.

\bibitem{WK}  Kenneth G. Wilson and J. Kogut, {\it The renormalization group and the $\epsilon$ expansion}, Phys. Repts. {\bf 12},  75 (1974).

\bibitem{1} M. E. Fisher and A. Aharony,
{\it Dipolar interactions at ferromagnetic critical points},
Phys. Rev. Lett. {\bf 30}, 559 (1973).


\bibitem{2} A. Aharony and M. E. Fisher,
{\it Critical behavior of magnets with dipolar interactions,
I. Renormalization group near four dimensions},
Phys. Rev. B {\bf  8}, 3323 (1973).

\bibitem{3} A. Aharony,
{\it Critical behavior of magnets with dipolar interactions,
II. Feynman graph expansion for ferromagnets near 4 dimensions},
Phys. Rev. B {\bf  8}, 3342 (1973).



\bibitem{4} A. Aharony,
{\it Critical behavior of magnets with dipolar interactions, III. Antiferromagnets},
Phys. Rev. B {\bf  8}, 3349(1973).

\bibitem{5} A. Aharony,
{\it Critical behavior of magnets with dipolar interactions, IV. Anisotropy},
Phys. Rev. B {\bf  8}, 3358 (1973).

\bibitem{6} A. Aharony,
{\it Critical behavior of magnets with dipolar interactions,
V. Uniaxial magnets in d dimensions},
Phys. Rev. B {\bf  8}, 3363 (1973). Erratum Phys Rev. B {\bf  9}, 3946 (1974).

\bibitem{cubic} A. Aharony, {\it Critical behavior of anisotropic cubic
  systems}, Phys. Rev. {\bf 8}, 4270 (1973).

  \bibitem{AADG} A. Aharony, {\it Dependence of universal critical behavior on symmetry and
  range of interaction}, in {\bf Phase Transitions and Critical
  Phenomena}, C. Domb and M. S. Green, eds., Vol. 6 (Academic Press, NY,
  1976), pp. 357-424.

  \bibitem{dipEOS} E. g., A. Aharony and A. D. Bruce, {\it Equation of state and scaling relations for isotropic ferromagnets with dipolar interactions}, Phys. Rev. B {\bf 10}, 2973 (1974).


  \bibitem{ahlers} A. Aharony and G. Ahlers, {\it Universal ratios among corrections to
  scaling amplitudes and effective critical exponents}, Phys. Rev. Lett.
  {\bf 44}, 782 (1980).

  \bibitem{GAlog} Guenter Ahlers, Avinoam Kornblit, and H. J. Guggenheim, {\it Logarithmic Corrections to the Landau Specific Heat near the Curie Temperature of the Dipolar Ising Ferromagnet LiTbF$^{}_4$}, Phys. Rev. Lett. {\bf 34}, 1227 (1975).




\bibitem{AB} E.g., A. D. Bruce and A. Aharony, {\it Coupled order parameters, symmetry
  breaking irrelevant scaling fields and tetracritical points}, Phys.
  Rev. B {\bf  11}, 478 (1975).

  \bibitem{KAM} A. Aharony, K. A. M\"{u}ller and W. Berlinger,
{\it Trigonal-to-tetragonal transition in stressed SrTiO$^{}_3$ : A realization of the 3-state Potts model},
Phys. Rev. Lett. {\bf 38}, 33 (1977). This paper was directly triggered by Ref. \cite{MFD}.

\bibitem{MFD} David Mukamel, Michael E. Fisher, and Eytan Domany, {\it Magnetization of Cubic Ferromagnets and the Three-Component Potts Model}, Phys. Rev. Lett. {\bf 37}, 565 (1976).

\bibitem{bruce2} The following 2 papers, describing theory and experiments, appeared back to back:  A. Aharony and A. D. Bruce, {\it Lifshitz point, critical and tricritical behavior in anisotropically-stressed Perovskites}, Phys. Rev. Lett.
 {\bf 42}, 462 (1979); J. Y. Buzaré, J. C. Fayet, W. Berlinger, and K. A. Müller, {\it
 Tricritical Behavior in Uniaxially Stressed RbCaF$^{}_3$},
Phys. Rev. Lett. {\bf 42}, 465 (1979).


\bibitem{KNF} J. M. Kosterlitz, D. R. Nelson  and M. E. Fisher,   {\it Bicritical and tetracritical points in anisotropic antiferromagnetic systems}, %\href{https://journals.aps.org/prb/abstract/10.1103/PhysRevB.13.412}{
    Phys. Rev. B {\bf 13}, 412 (1976).


\bibitem{DMF}  E. Domany, D. Mukamel, and M. E. Fisher, {\it Destruction of first-order transitions by symmetry-breaking fields}, %\href{https://journals.aps.org/prb/abstract/10.1103/PhysRevB.15.5432}
    Phys. Rev. B {\bf 15}, 5432 (1977).

\bibitem{brezin} E. Br\'{e}zin, D. J. Wallace, and Kenneth G. Wilson, {\it Feynman-Graph Expansion for the Equation of State near the Critical Point (Ising-like Case)}, Phys. Rev. Lett. {\bf 29}, 591 (1972).



\bibitem{sca1} M. E. Fisher and A. Aharony,
{\it Scaling function for critical scattering},
Phys. Rev. Lett. {\bf 31}, 1238 (1973).

\bibitem{sca2} M. E. Fisher and A. Aharony,
{\it Scaling function for two point correlations, I. Epsilon expansion near 4 dimensions},
Phys. Rev. B {\bf  10}, 2818 (1974).

\bibitem{katanin} A. Aharony, {\it Scaling function for two-point correlations. II. Expansion to order $1/n$}, Phys. Rev. B {\bf 10}, 2834 (1974).
Interestingly, these results were rediscovered independently by Andrey Katanin, in 2021! After our correspondence, which followed the ArXiv version of his paper, he found a misprint in my paper, and the corrected results appeared: A. Katanin, {\it Nonanalytic momentum dependence of spin susceptibility for Heisenberg magnets in the paramagnetic phase and its effect on critical exponents}, Phys. Rev. B {\bf 103}, 054415 (2021).

\bibitem{FL} M. E. Fisher and J. S. Langer, {\it Resistive anomalies at magnetic critical points}, Phys. Rev. Lett. {\bf 20}, 665 (1968).

\bibitem{privman} V. Privman, P. C. Hohenberg, and A. Aharony, {\it Universal critical-point
  amplitude relations}, in {\bf Phase Transitions and Critical Phenomena},
  C. Domb and J. L. Lebowitz, eds., Vol. 14 (Academic, NY, 1991), pp. 1-134,
  364-367.


\bibitem{AF1} A. Aharony and M. E. Fisher, {\it Universality in analytic corrections to
  scaling for planar Ising models}, Phys. Rev. Lett. {\bf 45}, 679
  (1980).
 \bibitem{AF2} A. Aharony and M. E. Fisher, {\it Nonlinear scaling fields and corrections
  to scaling near criticality}, Phys. Rev. B {\bf  27}, 4394 (1983).

  \bibitem{perk} Y. Chan, A.J. Guttmann, B.G. Nickel and J.H.H. Perk,
    {\it The Ising Susceptibility Scaling Function}, J. Stat. Phys. {\bf 145}  549 (2011) and references therein.

  \bibitem{gibbs} Michael E. Fisher, {\it Phases and Phase Diagrams -- Gibbs' Legacy Today}, Physica A {\bf 163}, 15 (1990).

  \bibitem{A5} A. Aharony, R. J. Birgeneau, A. Coniglio, M. A. Kastner, and H. E. Stanley,
{\it Magnetic phases and magnetic pairing in doped La$_2$CuO$_4$},
Phys. Rev. Lett. {\bf 60}, 1330 (1988).

\bibitem{bibl} Proverbs {\bf 13}, 20.

\bibitem{AH} A. Aharony and A. B. Harris,
{\it Absence of self-averaging and universal fluctuations in random systems near critical points},
Phys. Rev. Lett. {\bf 77}, 3700 (1996).

\bibitem{singh} P. B. Weichman and M.E.Fisher, 	{\it Helium in Vycor, constrained randomness and the Harris criterion}, Phys. Rev. B {\bf 34}, 7652 (1986); R .R. P. Singh and M. E.Fisher, {\it Random coupling crossover in Ising ferromagnets}, Phys. Rev. B {\bf 37}, 1980 (1988); R. P. Singh and Michael E. Fisher, {\it Disordered systems which escape the bound $\nu\geq 2/d$}, Phys. Rev. Lett. {\bf 60}, 548 (1988).

\bibitem{mef70} A. Aharony,
{\it Comment on ``Bicritical and tetracritical phenomena and scaling properties of the SO(5) theory''},
Phys. Rev. Lett. {\bf 88}, 059703 (2002); A. Aharony,
{\it Old and new results on multicritical points}, in Ref.~\cite{70}, dedicated to Fisher's 70th birthday, p. 659.


\bibitem{gefen1} Yuval Gefen,
Benoit B. Mandelbrot and
Amnon Aharony, {\it Critical Phenomena on Fractal Lattices}, Phys. Rev. Lett. {\bf 45}, 855 (1980).

\bibitem{gefen2} Yuval Gefen, Amnon Aharony,
Benoit B. Mandelbrot and Scott Kirkpatrick, {\it Solvable Fractal Family, and Its Possible Relation to the Backbone at Percolation}, Phys. Rev. Lett. {\bf 47}, 1771 (1981).

\bibitem{book} D. Stauffer and A. Aharony, {\bf
Introduction to Percolation Theory},
Taylor and Francis, London (1992); revised 2nd edition (1994).\footnote{ This book was written while Saddam Hussain sent missiles from Iraq to Israel. The universities were closed, and I had time to work in the bomb shelter. However, no journal editor allowed to acknowledge Hussain for helping my research -- so here it is. At that time Michael wrote to me about his experiences during the German Blitz bombing of London, and invited me and my family to come and stay in safe Maryland. We did not come, and stayed alive, but remain grateful for the kind intention.}

\bibitem{MEFperc} M. E. Fisher, {\it Critical probabilities for cluster size and percolation problems}, J Math. Phys. {\bf 2}, 620 (1960).

\bibitem{MEFfractals} M. E Fisher, {\it Fractal and nonfractal shapes in two-dimensional vesicles}, Physica D {\bf 38}, 112 (1989); Proc. Conf. “Fractals in Physics,” 1-4 October 1989, Vence, France, Eds. A. Aharony and J. Feder (North Holland Publ. Co., 1989).

\bibitem{6eps} E.g. L. T. Adzhemyan, E. V. Ivanova, M. V. Kompaniets, A. Kudlis, and A. I. Sokolov, {\it Six-loop
$\epsilon$ expansion study of three-dimensional $n-$vector model with cubic anisotropy}, Nucl. Phys. B {\bf 940}, 332 (2019) and references therein. See also chapter by Adzhemyan {\it et al.} in this book.

\bibitem{fixed-d} E.g., J. M. Carmona, A. Pelissato, and E. Vicari, {\it $N-$component Ginzburg-Landau Hamiltonians with cubic anisotropy: A six-loop study}, Phys. Rev. B {\bf 61}, 15136 (2000) and references therein.

\bibitem{boot} E.g., D. Poland, S. Rychkov, and A. Vichi,{\it  The conformal bootstrap: Theory, numerical techniques, and applications}, Rev. Mod. Phys. {\bf 91}, 015002 (2019); S. M. Chester, W. Landry, J. Liu, D. Poland, D. Simmons-Duffin, N. Su, and A. Vichi, {\it Bootstrapping Heisenberg magnets and their cubic anisotropy}, Phys. Rev. D {\bf 104}, 105013 (2021) and references therein. See also chapter by Poland and Simmons-Duffin in this book.

\bibitem{MC} E.g., M. Hasenbusch and E. Vicari, {\it Anisotropic perturbations in three-dimensional $O(N)-$symmetric vector models}, Phys. Rev. B {\bf 84}, 125136 (2011); Martin Hasenbusch, {\it Cubic fixed point in three dimensions: Monte Carlo simulations of the $\varphi^4$ model on the simple cubic lattice}, Phys. Rev. B {\bf 107}, 024409 (2023) and references therein.

\bibitem{AEK1} A. Aharony, O. Entin-Wohlman and A. Kudlis,
{\it Different critical behaviors in perovskites with a structural phase transition from cubic-to-trigonal and cubic-to-tetragonal symmetry},
Phys. Rev. B {\bf  105}, 104101 (2022).

\bibitem{AEK2} A. Aharony, O. Entin-Wohlman and A. Kudlis,
{\it Bi- and tetracritical phase diagrams in three dimensions},
Low Temperature Physics {\bf 48}, 483 (2022) [Fiz. Niz. Temp. (Kharkhov) {\bf 48}, 542 (2022)].

\bibitem{AE} A. Aharony and O. Entin-Wohlman,
{\it The puzzle of bicriticality in the XXZ antiferromagnet},
Phys. Rev. B {\bf 106}, 094424 (2022).


\bibitem{Rychkov} Slava Rychkov, {\it Numerical conformal bootstrap: targets}, Talk at the Simons Collaboration on the Nonperturbative Bootstrap Annual Meeting, Simons Foundation, New York, November 8-9 (2018). Recording is available:
https://www.simonsfoundation.org/event/simons-collaboration-on-the-nonperturbative-bootstrap-annual-meeting-2018/
Also, Aleix Gimenez-Grau, Yu Nakayama, Slava Rychkov,  unpublished.

\bibitem{brezin2} E. Br\'{e}zin and J. Zinn-Justin, {\it Critical behavior of uniaxial systems with strong dipolar interactions}, Phys. Rev. B {\bf  13}, 251 (1976).




\bibitem{AIM} Amnon Aharony, Yoseph Imry, and Shang-keng Ma, {\it Lowering of Dimensionality in Phase Transitions with Random Fields}, Phys. Rev. Lett. {\bf 37}, 1364 (1976).

\bibitem{parisi} G. Parisi and N. Sourlas, {\it Random Magnetic Fields, Supersymmetry, and Negative Dimensions},
Phys. Rev. Lett. {\bf 43}, 744 (1979).

\bibitem{amp} A. Aharony and B. I. Halperin, {\it Exact relations among amplitudes at
  critical points of marginal dimensionality}, Phys. Rev. Lett. {\bf 35},
  1308 (1975).


\bibitem{new} A. Aharony, {\it New singularities in the critical behavior of
  random Ising models at marginal dimensionalities}, Phys. Rev.
  B {\bf  13}, 2092 (1976).



\bibitem{expdd} A. Ferdinand, A.-C. Probst, A. Michels, R. Birringer and S. N. Kaul, {\it Critical behaviour of nanocrystalline gadolinium: evidence for random uniaxial dipolar universality class}, J. Phys.: Condens. Matter {\bf 26}, 056003 (2014).


\bibitem{shechter} Tomer Dollberg, Juan Carlos Andresen, and Moshe Schechter, {\it Effect of intrinsic quantum fluctuations on the phase diagram of anisotropic dipolar magnets}, Phys. Rev. B {\bf 105}, L180413 (2022).
\bibitem{elliot} R. J. Elliott, P. Pfeuty and C. Wood, {\it Ising model with a transverse field}, Phys. Rev. Lett. {\bf 25}, 443 (1970).

    \bibitem{suzuki} M. Suzuki, {\it Relationship between $d-$dimensioal quantal spin systems and $(d+1)-$dimensional Ising systems}, Prog. Theor. Phys. {\bf 56}, 1454 (1976).
    \bibitem{sachdev} Subir Sachdev, {\bf Quantum Phase Transitions}, Cambridge University Press (2011).

    \bibitem{Qshift} E.g., David Andelman and Amnon Aharony, {\it Critical behavior with axially correlated random bonds},
Phys. Rev. B {\bf 31}, 4305 (1985) and references therein.

    \bibitem{nat} Cubic symmetry is reflected by terms like $hq^2_\alpha\delta^{}_{\alpha\beta}$ in Eq. (\ref{Uab}).~\cite{2} In parallel to our work, Thomas Nattermann sat isolated in East Germany, and impressively found indepedently many of our dipolar results (produced in the supporting Cornell atmosphere). In particular, he concentrated on these cubic terms: e.g.,  T. Nattermann and S. Trimper, {\it Critical behaviour and cubic anisotropy}, J. Phys. A: Math. Gen., {\bf 8}, 2000 (1975). Here I ignore these terms for simplicity. They (as well as the terms with $q^\alpha q^\beta$ also seem to be irrelevant, but only at order $\epsilon^2$. In those years Nattermann and I corresponded a lot by snail mail (this was before the electronic mail!), and met many times after he escaped from East Germany and joined the university of K\"{o}ln in West Germany.


    \bibitem{pfeuty} M. E. Fisher and P. Pfeuty, {\it Critical Behavior of the Anisotropic $n-$Vector Model}, Phys. Rev. B {\bf 6}, 1889 (1972).


\bibitem{ND} David R. Nelson and Eytan Domany, {\it Equations of state for bicritical points. I. Calculations in the disordered phase}, Phys. Rev. B {\bf 13}, 236 (1976).


\bibitem{reid} K. Reid, Y. Millev, M. F\"{a}hnle and H. Kronm\"{u}ller, {\it Phase transitions in ferromagnets with dipolar interactions and uniaxial anisotropy}, Phys. Rev. B {\bf 51}, 15229 (1995).

\bibitem{kaul} S.N. Kaul,  {\it Critical Behaviour of Heisenberg Ferromagnets with Dipolar Interactions and Uniaxial Anisotropy}, Lect. Notes Phys. {\bf 678}, 11 (2005).










%\bibitem{afm} If the exchange is antiferromagnetic, we need to double the unit cell, define two orderparameters (the sum and difference of neighboring spins in the doubled  unit ell, and then eliminate the antiferroamgnetic order parameter (the above difference) from the partition function. See e.g., D. R. Nelson and M. E. Fisher, {\it Renormalization-group analysis of metamagnetic tricritical behavior}, Phys. Rev. B {\bf 11}, 1030(1975). For simplicity, we ignore this possibility.



\bibitem{GrL} S. Grinstein and A. Luther, {\it Application of the reormalization group to phase transitions in disordered systems}, Phys. Rev. B {\bf 13}, 1329 (1976).


\bibitem{AAran} A. Aharony, {\it Critical properties of random and constrained dipolar magnets}, Phys. Rev. B {\bf 12}, 1049 (1975).

    \bibitem{ma} A. Aharony, Y. Imry and S. Ma, {\it  Comments on the critical behavior of random systems}, Phys. Rev. B {\bf 13}, 466 (1976). This reference averaged over the local transition temperature, while here we average over the site occupation.

%\bibitem{avg} The average $\langle \exp(-\beta{\cal H}^{}_n)\rangle$ requires~\cite{ma}
%$\langle p^{}_{\bf R}\rangle=1-x,\ \langle p^{}_{\bf R} p^{}_{\bf R'}\rangle=(1-x)^2+x(1-x)\delta({\bf R}-{\bf R'})$.
%For ${\bf R}\ne{\bf R'}$ and ${\bf R''}\ne{\bf R'''}$ one has
%$\langle p^{}_{\bf R} p^{}_{\bf R'}p^{}_{\bf R''}\rangle=(1-x)^3+x(1-x)^2[\delta({\bf R}-{\bf R''})+\delta({\bf R'}-{\bf R''})]$,
%$\langle p^{}_{\bf R} p^{}_{\bf R'}p^{}_{\bf R''}p^{}_{\bf R'''}\rangle=(1-x)^4+x(1-x)^3[\delta({\bf R}-{\bf R''})+\delta({\bf R'}-{\bf R''})+\delta({\bf R}-{\bf R'''})+\delta({\bf R'}-{\bf R'''})]
%+x^2(1-x)^2[\delta({\bf R}-{\bf R''})\delta({\bf  R'}-{\bf R'''})+\delta({\bf R'}-{\bf R''})\delta({\bf R}-{\bf R'''})]$.


\bibitem{AASSC} A. Aharony, {\it Absence of ferromagnetic long range order in random isotropic dipolar magnets and in similar systems}, Solid State Comm. {\bf 28}, 667 (1978).

\bibitem{schra} M. Schechter, LiHo$^{}_x$Y$^{}_{1-x}$F$^{}_4$ {\it as a random-field Ising ferromagnet}, Phys Rev. B {\bf 77}, 020401(R) (2008).

\bibitem{aepli} D. M. Silevitch, D. Bitko, J. Brooke, S. Ghosh, G. Aeppli and T. F. Rosenbaum, {\it A ferromagnet in a continuously tunable random field}, Nature {\bf 448}, 567– (2007).

\bibitem{liu} K S. Liu and M. E. Fisher, {\it Ouantum lattice gas and the existence of a supersolid}, J. Low Temp. Phys. {\bf 10}, 655 (1973).


\bibitem{KW} I. J. Ketley and D. J. Wallace, {\it A modified epsilon expansion for a Hamiltonian with cubic point-group symmetry},  J. Phys. A: Math. Nucl. Gen. {\bf 6} 1667 (1973).

    \bibitem{hu} Xiao Hu, {\it Bicritical and Tetracritical Phenomena and Scaling Properties of the SO(5) Theory},
Phys. Rev. Lett. {\bf 87}, 057004 (2001).

\bibitem{zhang}  E. Demler, W. Hanke, and S.-C. Zhang, {\it $SO(5)$ theory of antiferromagnetism and superconductivity}, Rev. Mod. Phys. {\bf 76}, 909 (2004).

    \bibitem{AEK3} A. Kudlis, A. Aharony and O. Entin-Wohlman, {\it Effective exponents near bicritical points}, ArXiv:2304.08265. To appear in EPJ special edition on Non-Equilibrium Quantum Physics, Many Body Systems, and Foundations of Quantum Mechanics.

\bibitem{wegner} F. J. Wegner, {\it Critical Exponents in Isotropic Spin Systems}, %\href{https://journals.aps.org/prb/abstract/10.1103/PhysRevB.6.1891}{
     Phys. Rev. {\bf B 6}, 1891 (1972). %See also F. J. Wegner, {\it The critical state, General Aspects},  in Ref. \cite{DG}, p. 7.

   \bibitem{zan} A. Codello, M. Safari, G. P. Vacca, and O. Zanusso, {\it Critical models with $n\leqq 4$
 scalars in $d=4-\epsilon$}, %\href{https://journals.aps.org/prd/abstract/10.1103/PhysRevD.102.065017}{
 Phys. Rev. D {\bf 102}, 065017 (2020) and references therein.


\bibitem{vicari} P. Calabrese, A. Pelissetto, and E. Vicari, {\it Multicritical phenomena in $O(n^{}_1)\bigoplus O(n^{}_2)-$symmetric theories}, %\href{https://journals.aps.org/prb/abstract/10.1103/PhysRevB.67.054505}{
    Phys. Rev. B {\bf 67}, 054505 (2003).


\bibitem{FA} S. Fishman and A. Aharony, {\it Random field effects in disordered
  anisotropic antiferromagnets}, J. Phys. C: Solid State Phys. {\bf 12},
  L729 (1979).

  \bibitem{cardy} J. L. Cardy, {\it Random-field effects in site disordered Ising antiferromagnets}, Phys. Rev. B {\bf 29}, 505 (1984).

 \bibitem{deG} P. G. de Gennes, {\it Liquid-liquid demixing inside a rigid network. Qualitative features}, J. Phys. Chem. {\bf 88}, 6469 (1984).

\bibitem{rohrer} H. Rohrer, A. Aharony and S. Fishman, {\it Critical and multicritical properties of random antiferromagnets}, J. Mag. Mag. Mat. {\bf 15-18}, 396 (1980).

\bibitem{expts} E.g., J. P. Hill, T. R. Thurston, R. W. Erwin, M. J. Ramstad, and R. J. Birgeneau, {\it Transition to long-range order in the three-dimensional random-field Ising model}, Phys. Rev. Lett. {\bf 66}, 3281 (1991).

\bibitem{cowley} R. A. Cowley, A. Aharony, R. J. Birgeneau, R. A. Pelcovits, G. Shirane and T. R. Thurston, {\it The bicritical phase diagram of two-dimensional antiferromagnets with and without random fields}, Zeitschrift f\"{u}r Physik B Condensed Matter {\bf 93}, 5 (1993);
R. J. Birgeneau, A. Aharony, R. A. Cowley, J. P. Hill, R. A. Pelcovits, G. Shirane and T. R. Thurston, {\it Effects of random fields on bicritical phase diagrams in two and three dimensions}, in Ref.~\cite{70}, dedicated to Fisher's 70th birthday, p. 58.


\bibitem{lacour} P. Lacour-Gayet et G. Toulouse, {\it Ideal Bose Einstein condensation and disorder effects}, J. Phys. France {\bf 35}, 425 (1974).

\bibitem{IM} Y. Imry and S.-k. Ma, {\it Random-fields instability of the ordered state of continuous symmetry}, Phys. Rev. Lett. {\bf 35}, 1399 (1975).

%\bibitem{AIM} A. Aharony, Y. Imry and S. Ma, {\it Lowering of dimensionality in phase transitions with random fields}, Phys. Rev. Lett. {\bf 37}, 1364 (1976).

\bibitem{nowak} U. Nowak {\it et al.}, {\it Dyamic state model for exchage bias. I. Thory}, Phys. Rev. B {\bf 66}, 014430 (2002).

\bibitem{rjb} R. J. Birgeeau, {\it Random fields ad phase transitions in model magnetic systems}, J. Mag. Mag. Mat. {\bf 177-178}, 1 (1998).


\bibitem{young} See also P. A. Young, {\it On the lowering of dimensionality in phase trasitions with random fields}, J Physics C {\bf 10}, L257 (1977).

%\bibitem{PS} G. Parisi and  N. Sourlas, {\it Random Magnetic Fields, Supersymmetry, and Negative Dimensions}, Phys. Rev. Lett. {\bf 43}, 744 (1979).


\bibitem{imbrie} John Z. Imbrie, {\it Lower Critical Dimension of the Random-Field Ising Model}, Phys. Rev. Lett. {\bf 53}, 1747 (1984).

\bibitem{brickmont} J. Brickmont and A. Kupiainen, {\it Lower critical dimension for the random-field Ising model}, Phys. Rev. Lett. {\bf 59}, 1829 (1987).



\bibitem{AARF} A. Aharony, {\it Tricritical points in systems with random fields}, Phys. Rev. B {\bf 18}, 3318 (1978).

    \bibitem{andelman} D. Andelman, {\it First- and second-order phase trasitions with random fields at low temperatues}, Phys. Rev. B {\bf 27}, 3079 (1983).

\bibitem{shapirs} Yonathan Shapir and Amnon Aharony,  {\it High-temperature series and exact relations for the Ising model in a random field},  J. Phys. C: Solid State Phys. {\bf 15}, 1361 (1982).

\bibitem{shapir2}  Y. Shapir and A. Aharony, {\it Rigorous relations for the cross-over behaviour
due to random perturbations},   J. Phys. C: Solid State Phys. {\bf 14},  L905 (1981).





\bibitem{gofman} Misha Gofman, Joan Adler, Amnon Aharony, A. B. Harris, and Moshe Schwartz,
            {\it Evidence for two exponent scaling in the random field Ising model}, Phys. Rev. Lett. {\bf 71}, 1569 (1993); {\it Critical behavior of the random-field Ising model}, Phys. Rev. B {\bf 53}, 6362 (1996).

\bibitem{MEF1999} M. E. Fisher, {\it Some views from forty years as a statistical mechanician}, Proc. STATPHYS20,  Physica A {\bf 263}, 554 (1999).

\bibitem{grinstein} e.g., G. Grinstein and S.-k. Ma, {\it Roughening and lower critical dimension in the random-field Ising model}, Phys. Rev. Lett. {\bf 49}, 685 (1982); D. S. Fisher, {\it Interface fluctuations in disordered systems: $5-\epsilon$ expansion and failure of dimensional reduction}, Phys. Rev. Lett. {\bf 56}, 1964 (1986).

\bibitem{bray} A. J. Bray and M. A. Moore, {\it Scaling theory of the random-field Ising model}, J. Phys. C:  Solid State Phys. {\bf 18}, L927 (1985).


\bibitem{1st-order} M. E. Fisher and A. N. Berker, {\it Scaling for first-order transitions in thermodynamic and finite systems}, Phys. Rev. B {\bf 26}, 2507 (1982).
    
    \bibitem{swift} M. R. Swift, A. J. Bray, A. Marita, M. Cieplak and J. R. Baavar, {\it Scaling of the random-field Ising model at zero temperature}, EuroPhys. Lett. {\bf 38}, 273 (1997).
        
        \bibitem{sourlas} N. Sourlas, {\it Universality in random systems: the case of the 3D random field Ising model}, Computer Phys. Comm. {\bf 121-122}, 183 (1999).
        
        \bibitem{self} A. Aharony and A. B. Harris, {\it Absece of self-averagig and uiversal fluctuatios i random systems near critical poits}, Phys. Rev. Lett. {\bf 77}, 3700 (1996).
            
            \bibitem{ParSour} G. Parisi and N. Sourlas, {\it Scale invariance in disordered systems: The example of the  random-field Ising model}, Phys. Rev. Lett. {\bf 89}, 257204 (2002).

\bibitem{RFMC} Nikolaos G. Fytas and V\'{\i}ctor Mart\'{\i}n-Mayor, {\it Universality in the Three-Dimensional Random-Field Ising Model}, Phys. Rev. Lett. {\bf 110}, 227201 (2013); Nikolaos G. Fytas, V\'{\i}ctor Martín-Mayor, Marco Picco, and Nicolas Sourlas, {\it Phase Transitions in Disordered Systems: The Example of the Random-Field Ising Model in Four Dimensions}, Phys. Rev. Lett. {\bf 116}, 227201 (2016); Nikolaos G. Fytas, V\'{\i}ctor Martín-Mayor, Giorgio Parisi, Marco Picco, and Nicolas Sourlas, {\it Evidence for Supersymmetry in the Random-Field Ising Model at $D=5$}, Phys. Rev. Lett. {\bf 122}, 240603 (2019).

\bibitem{nikos} Nikolaos G. Fytas and V\'{\i}ctor Mart\'{\i}n-Mayor, {\it Efficient numerical methods for the random-field Ising model:
Finite-size scaling, reweighting extrapolation, and computation of response functions}, Phys. Rev. E {\bf 93}, 063308 (2016).
   
\bibitem{IW} Y. Imry and  M. Wortis, {\it Influence of quenched impurities on first-order phase transitions}, Phys. Rev. B {\bf 19}, 3580 (1979).

\bibitem{Aizen} M. Aizenman and J. Wehr, {\it Rounding of first-order phase trasitions in systems with quenched disorder}, Phys. Rev. Lett. {\bf 62}, 2503 (1989).
\bibitem{Berker} K. Hui and A. N. Berker, {\it Random-field mechanism in random-bond multicritical systems}, Phys. Rev. Lett. {\bf 62}, 2507 (1989).


\bibitem{kumar} M. Kumar, V. Banerjee and S. Puri, {\it Random field Ising model in a uniform magnetic field: Ground states, pinned clusters and scaling laws}, Eur. Phys. J. E {\bf 40}, 96 (2017).
    
    \bibitem{specific} Similar apparent first order trasitions for sigle realizations were also indicated by L. Hern\'{a}ndez ad H. Ceva, {\it Wang-Landau study of the critical behavior of the bimodal 3D random field Ising model}, Physica A {\bf 387}, 2793 (2008)   and by N. G. Fytas and A. Malakis, {\it Phase diagram of the 3D bimodal random-field Ising model}, Eur. Phys. J. B {\bf 61}, 111 (2008).

\bibitem{br-deDom} E. Br\'{e}zin and C. De Dominicis, {\it New phenomena in the random field Ising model}, Europhys. Lett. {\bf 44}, 13 (1998).

\bibitem{parisi11} In fact, such a possibility was already mentioned by G. Parisi, {\it An introduction to the statistical mechanics of amorphous systems}, in {\bf Recent advances in field theory and statistical mechanics}, ed. J.-B. Zuber and R. Stora (Elsevier, 1984).

 \bibitem{TT}  G. Tarjus and M. Tissier, {\it Random-field Ising model and $O(N)$ models: theoretical description through the functional renormalization group}, Eur. Phys. J. B {\bf 93}, 50 (2020) and references therein.
  
  
  %M. Tissier and G. Tarjus, {\it Nonperturbative functional renormalization group for random-field models: the way out o dimesioal reduction}, Phys. Rev. Lett. {\bf 93}, 267008 (2004); {\it Supersymmetry and its spontaneous breaking in the random field Ising model}, {\it ibid.} {\bf 107}, 041601 (2011); G. Tarjus and M. Tissier, {\it Avalanches and perturbation theory in the random-field Ising model}, J. Stat. Mech. {\bf 2016}, 023207 (2016).

\bibitem{RFslava} Apratim Kaviraj, Slava Rychkov, and Emilio Trevisani, {\it Parisi-Sourlas Supersymmetry in Random Field Models}, Phys. Rev. Lett. {\bf 129}, 045701 (2022); {\it Random field Ising model and Parisi-Sourlas supersymmetry I. Supersymmetric CFT}, JHEP {\bf 04}, 090 (2020);  {\it Random field Ising model and Parisi-Sourlas supersymmetry. Part II. Renormalization group}, arXiv:2009.10087; S. Rychkov, {\it Four lectures on the random field Ising model, Parisi-Sourlas supersymmetry and dimensional reduction}, arXiv:2303.09654. These lecture notes contain many relevant references. 

\bibitem{maritan} A. Maritan {\it et al.}, {\it Orderig and phase trasitions in random-field Ising model}, Phys. Rev. Lett. {\bf 67}, 1821  (1991).
\end{thebibliography}
\end{document}